\documentclass[aps,pra,twocolumn]{revtex4}
\usepackage{amssymb,amsbsy,graphicx,times,subfigure,marvosym,color,bm,multirow}
\usepackage[latin1]{inputenc}
\begin{document}

%%%%%%%%%%%%%%%%%%%%%%%%%%%%%%%%%%%%%%%%%%%%%%%%

\title{Probing topological order with R\'{e}nyi entropy}
\author{G\'{a}bor B. Hal\'{a}sz$^{1,2}$}
\author{Alioscia Hamma$^{1,3}$}
\address{$^1$Perimeter Institute for Theoretical Physics, 31 Caroline Street North, Waterloo, Ontario, Canada N2L 2Y5 \\
$^2$Theoretical Physics, Oxford University, 1 Keble Road, Oxford OX1 3NP, United Kingdom \\
$^3$Center for Quantum Information, Institute for Interdisciplinary
Information Sciences, Tsinghua University, Beijing 100084, People's
Republic of China}

%%%%%%%%%%%%%%%%%%%%%%%%%%%%%%%%%%%%%%%%%%%%%%%%

\begin{abstract}

We present an analytical study of the quantum phase transition
between the topologically ordered toric-code-model ground state and
the disordered spin-polarized state. The phase transition is induced
by applying an external magnetic field, and the variation in
topological order is detected via two non-local quantities: the
Wilson loop and the topological R\'{e}nyi entropy of order $2$. By
exploiting an equivalence with the transverse-field Ising model and
considering two different variants of the problem, we investigate
the field dependence of these quantities by means of an exact
treatment in the exactly solvable variant and complementary
perturbation theories around the limits of zero and infinite fields
in both variants. We find strong evidence that the phase transition
point between topological order and disorder is marked by a
discontinuity in the topological R\'{e}nyi entropy and that the two
phases around the phase transition point are characterized by its
different constant values. Our results therefore indicate that the
topological R\'{e}nyi entropy is a proper topological invariant: its
allowed values are discrete and can be used to distinguish between
different phases of matter.

\end{abstract}

%%%%%%%%%%%%%%%%%%%%%%%%%%%%%%%%%%%%%%%%%%%%%%%%

\maketitle

%%%%%%%%%%%%%%%%%%%%%%%%%%%%%%%%%%%%%%%%%%%%%%%%%

\section{Introduction} \label{sec-int}

There are states in quantum many-body physics that cannot be
described in terms of local order parameters and the Landau paradigm
of spontaneous symmetry breaking. These states exhibit a subtler
kind of order called topological order \cite{Wen}. Topologically
ordered states include fractional quantum Hall liquids \cite{FQHE}
and quantum spin liquids \cite{Isakov}, which are at the forefront
of research in condensed matter theory. Moreover, such states are of
great interest in the field of quantum computation because one can
encode quantum information in the topological degrees of freedom and
this way of encoding is intrinsically robust against decoherence
\cite{Kitaev, QC}.

Since topologically ordered states cannot be characterized by local
order parameters, there has been an intense effort to find non-local
quantities that can detect topological order in a wavefunction. A
series of papers suggested that topological order can be detected
through a component of quantum entanglement that contains a
topological constraint. This constraint manifests itself as a
universal negative correction to the boundary law for the
entanglement entropy: the so-called topological entropy
\cite{Hamma-1, TE, Flammia, Kim}. Recent works have showed that this
component of quantum entanglement is indeed long ranged and so it
cannot be destroyed by time evolution with a local Hamiltonian.
Equivalently, the components corresponding to the long-range
entanglement and the usual short-range entanglement are
adiabatically disconnected \cite{Chen}.

These results suggest that the topological entropy is to some extent
a non-local order parameter for topologically ordered phases. To
make a more precise statement about the extent of its applicability,
one needs to investigate its robustness against perturbations. If
the topological entropy is to detect topologically ordered phases,
it needs to be non-zero within all such phases. In other words, it
should only vanish at quantum phase transitions to disordered
phases. If it is to distinguish different topologically ordered
phases from each other, it needs to be constant within each phase.
In other words, it should only change at quantum phase transitions
\cite{Hamma-2}.

Recent numerical studies on small systems have found evidence that
the topological entropy takes discrete values \cite{Isakov} and only
changes at quantum phase transitions \cite{Hamma-3}. On the other
hand, analytic corrections to the topological entropy are extremely
hard to obtain because one needs to consider the entanglement
entropy of a many-body wavefunction \cite{Amico}. Such a treatment
for the topological entropy in the case of a finite correlation
length can be found in Ref. \cite{Papanikolaou}. Remarkably, it has
been recently shown for an exactly solvable two-phase system that
the topological entropy is constant within the entire topologically
ordered phase \cite{Castelnovo}.

Given the difficulties, it is important to find other entropic
quantities that possess a topological component capable of detecting
topological order. One potential candidate is the R\'{e}nyi entropy
of order $\alpha$, which is a generalization of the usual (von
Neumann) entanglement entropy. It is important that the R\'{e}nyi
entropy coincides with the entanglement entropy in the special case
of $\alpha = 1$. It has also been shown that the ground-state
R\'{e}nyi entropies of different order $\alpha$ all contain the same
topological component at the fixed points of non-chiral phases
\cite{Flammia}. Such phases typically appear in string-net models
\cite{Levin} and quantum double models \cite{Buerschaper}.

In this paper, we consider the R\'{e}nyi entropy of order $2$ and
argue that it is a good probe of topological order because its
topological component can only change at quantum phase transitions.
In particular, we apply the concept of the topological R\'{e}nyi
entropy to the toric-code model (TCM) \cite{Kitaev} in the presence
of an external magnetic field. This model is, to paraphrase
Goldenfeld \cite{Goldenfeld}, the \emph{Drosophila} of topological
order. Although it is a simple toy model, it contains all the
elements that make topological order interesting: there is no local
order parameter, there is a topology-dependent ground-state
degeneracy that is robust against local perturbations, and there are
excitations with anyonic particle statistics. Indeed, the TCM is
another beautiful example of the crucial role played by toy models
in statistical mechanics.

To show that the topological R\'{e}nyi entropy is a good probe of
topological order, we demonstrate that the disordered and the
topologically ordered phases of the TCM with external magnetic field
are characterized by its distinct values. We also study the Wilson
loop as a probe of both topological order and gauge structure.
Concentrating on two different variants of the problem, we establish
an exact treatment in the computationally simpler (exactly solvable)
variant and supplement it with perturbation theories in both
variants. The results obtained with the two methods for the two
probing quantities in the two variants are highly consistent with
each other.

\section{General formalism} \label{sec-gen}

\begin{figure}[h!]
\centering
\includegraphics[width=4.0cm]{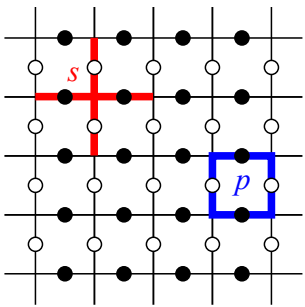}
\caption{(Color online) Illustration of the square lattice with the
physical spins located at the horizontal (black circles) and the
vertical (white circles) edges. Examples of a star (red cross
labeled $s$) and a plaquette (blue square labeled $p$) are included.
\label{fig-1}}
\end{figure}

We consider the TCM with an external magnetic field in the $+z$
direction. The system is an $N \times N$ square lattice with
periodic boundary conditions, and $2N^2$ spins are located at the
edges of the lattice \cite{Kitaev}. In general, the spins on the
horizontal ($h$) and the vertical ($v$) edges experience different
magnetic fields: $\lambda$ on the horizontal and $\kappa \lambda$ on
the vertical edges ($\kappa > 0$). The Hamiltonian of the system
takes the form
\begin{eqnarray}
\hat{H} &=& - \sum_s \hat{A}_s - \sum_p \hat{B}_p - \lambda \sum_{i
\in h} \hat{\sigma}_i^z - \kappa \lambda \sum_{i \in v}
\hat{\sigma}_i^z, \nonumber \\
&& \hat{A}_s \equiv \prod_{i \in s} \hat{\sigma}_i^x, \qquad
\hat{B}_p \equiv \prod_{i \in p} \hat{\sigma}_i^z,
\label{eq-gen-H-old}
\end{eqnarray}
where the indices $s$ and $p$ refer to stars and plaquettes on the
lattice containing four spins each. For an illustration of this, see
Fig. \ref{fig-1}. Note that the four sums in Eq.
(\ref{eq-gen-H-old}) all contain $N^2$ terms, and that only $N^2 -
1$ star (plaquette) operators are independent because $\prod_s
\hat{A}_s = \prod_p \hat{B}_p = 1$.

The TCM with zero external field ($\lambda = 0$) is exactly solvable
because the stars $\hat{A}_s$ and the plaquettes $\hat{B}_p$ all
commute with each other. The ground state is fourfold degenerate:
there are four linearly independent states with $A_s \equiv \langle
\hat{A}_s \rangle = +1$ and $B_p \equiv \langle \hat{B}_p \rangle =
+1$ for all $s$ and $p$. These degenerate ground states are
distinguished by the topological quantum numbers $Z_1 = \pm 1$ and
$Z_2 = \pm 1$, which are expectation values for products of
$\hat{\sigma}_i^z$ operators along horizontal and vertical strings
going round the lattice. The ground state with $Z_1 = Z_2 = +1$ can
be written as
\begin{equation}
| 0 \rangle = \mathcal{N} \prod_{s} \left( 1 + \hat{A}_s \right) |
\Uparrow \, \rangle, \label{eq-gen-gs}
\end{equation}
where $\mathcal{N} = 1 / \sqrt{2^{N^2+1}}$ is a normalization
constant, and $| \Uparrow \, \rangle$ denotes the completely
polarized state with all spins pointing in the $+z$ direction
($\sigma_i^z = +1$ for all $i$).

The TCM with finite external field ($\lambda > 0$) is not exactly
solvable because the magnetic fields $\hat{\sigma}_i^z$ do not
commute with the stars $\hat{A}_s$. On the other hand, they commute
with the plaquettes $\hat{B}_p$ and the topological operators
$\hat{Z}_{1,2}$, therefore there are four independent lowest-energy
sectors with $Z_{1,2} = \pm 1$ and $B_p = +1$ ($\forall p$). In the
rest of the paper, we consider the lowest-energy eigenstate $|
\Omega (\lambda) \rangle$ within the $Z_1 = Z_2 = +1$ sector. This
state becomes $| \Uparrow \, \rangle$ in the limit of $\lambda
\rightarrow \infty$ and $| 0 \rangle$ in the limit of $\lambda = 0$.
Between the two limits, numerical studies reveal a quantum phase
transition at a critical magnetic field $\lambda = \lambda_C$
\cite{Hamma-3, Trebst}. Since $| \Omega (\lambda) \rangle$ is a
ground state at the fixed points of both limiting phases, the
adiabatic theorem guarantees that it is the unique ground state in
the disordered phase at $\lambda > \lambda_C$ and one of the four
degenerate ground states in the topologically ordered phase at
$\lambda < \lambda_C$.

If we only consider the states with $Z_1 = Z_2 = +1$ and $B_p = +1$
($\forall p$), the dimension of the effective Hilbert space is
reduced from $2^{2N^2}$ to $2^{N^2-1}$. The states within this
reduced Hilbert space can be written as superpositions of loop
configurations on the dual lattice: in each loop configuration, the
spins on the loops have $\sigma_i^z = -1$ and the remaining spins
have $\sigma_i^z = +1$. This implies that the reduced model is
equivalent to a $\mathbb{Z}_2$ lattice gauge theory, and the phase
transition at $\lambda = \lambda_C$ corresponds to a
confinement-deconfinement transition \cite{Hamma-3, Trebst}.
Furthermore, since each loop configuration can be characterized by
the values of the $N^2 - 1$ independent stars $A_s = \pm 1$, it is
convenient to introduce a corresponding representation in which
quasi-spins $A_s$ are located at the stars \cite{Dusuel}. This
quasi-spin representation is particularly useful in the $\lambda \ll
1$ limit because $| \Omega (\lambda) \rangle$ is then close to $| 0
\rangle$ which is a product state with $A_s = +1$ ($\forall s$). Up
to an irrelevant additive constant, the Hamiltonian in Eq.
(\ref{eq-gen-H-old}) becomes
\begin{equation}
\hat{H} = - \sum_s \hat{A}_s^z - \lambda \sum_{\langle s,s' \rangle
\in h} \hat{A}_s^x \hat{A}_{s'}^x - \kappa \lambda \sum_{\langle
s,s' \rangle \in v} \hat{A}_s^x \hat{A}_{s'}^x, \label{eq-gen-H-new}
\end{equation}
where $\langle s,s' \rangle$ means that the summation is over
horizontal and vertical edges between nearest-neighbor stars $s$ and
$s'$. Note that $\hat{A}_s^z \equiv \hat{A}_s$ measures and
$\hat{A}_s^x$ switches the quantum number $A_s$, therefore the
quasi-spin operators $\hat{A}_s^z$ and $\hat{A}_s^x$ satisfy the
standard spin commutation relations. In the quasi-spin
representation of Eq. (\ref{eq-gen-H-new}), the TCM with external
magnetic field is equivalent to a two-dimensional (2D)
transverse-field Ising model (TFIM) in which the coupling strengths
on the horizontal and the vertical edges are different in general.

\section{Measures of topological order} \label{sec-ord}

We aim to describe how the topological order in the ground state $|
\Omega (\lambda) \rangle$ changes as a function of $\lambda$ between
the topologically ordered limit at $\lambda = 0$ and the disordered
limit at $\lambda \rightarrow \infty$. To quantify topological order
in an analytically tractable manner, we consider two measures: the
Wilson loop and the topological R\'{e}nyi entropy.

\subsection{General properties} \label{sec-ord-gen}

\begin{figure}[h!]
\centering
\includegraphics[width=4.1cm]{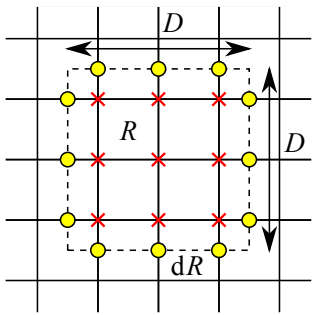}
\caption{(Color online) Illustration of the Wilson loop for a square
region with $D = 3$. The region $R$ contains $D^2$ stars (red
crosses) and the boundary $\partial R$ (dashed line) contains $4D$
spins (yellow circles). \label{fig-2}}
\end{figure}

The Wilson loop for a region $R$ on the dual lattice is defined as
the expectation value $W_R$ of the operator
\begin{equation}
\hat{W}_R \equiv \prod_{i \in \partial R} \hat{\sigma}_i^x =
\prod_{s \in R} \hat{A}_s^z, \label{eq-ord-gen-W}
\end{equation}
where $\partial R$ denotes the boundary of $R$. If the region $R$ is
macroscopic with linear dimension $D \gg 1$, the Wilson loop follows
a perimeter law $W_R \propto \exp (-\beta D)$ in the presence of
topological order and an area law $W_R \propto \exp (-\beta D^2)$ in
the absence of topological order \cite{Hamma-3}. In this paper, we
assume that the region $R$ is a $D \times D$ square (see Fig.
\ref{fig-2}).

\begin{figure}[h!]
\centering
\includegraphics[width=8.5cm]{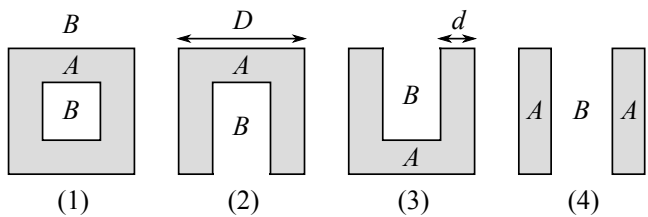}
\caption{Illustration of the subsystems $A$ and $B$ in the four
cases $(m)$ that are used to calculate the topological R\'{e}nyi
entropy. Each subsystem $A$ has extension $D$ and thickness $d$ with
$D > d \gg 1$. \label{fig-3}}
\end{figure}

The topological R\'{e}nyi entropy is based on the paradigm of
quantum entanglement. The R\'{e}nyi entropy of order $\alpha$
between two complementary subsystems $A$ and $B \equiv \overline{A}$
reads
\begin{equation}
S_{\alpha}^{AB} \equiv \frac{1} {1-\alpha} \log_2 \mathrm{Tr} \left[
\hat{\rho}_A^{\alpha} \right] = \frac{1} {1-\alpha} \log_2
\mathrm{Tr} \left[ \hat{\rho}_B^{\alpha} \right],
\label{eq-ord-gen-S-renyi}
\end{equation}
where $\hat{\rho}_A$ and $\hat{\rho}_B$ are the reduced density
operators for $A$ and $B$. The topological contribution to the
R\'{e}nyi entropy can be extracted by taking a suitable linear
combination of R\'{e}nyi entropies that are calculated for different
choices of the subsystems $A$ and $B$ \cite{TE}. In fact, the
standard definition for the topological R\'{e}nyi entropy of order
$\alpha$ is
\begin{equation}
S_{\alpha}^T \equiv - S_{\alpha}^{(1)} + S_{\alpha}^{(2)} +
S_{\alpha}^{(3)} - S_{\alpha}^{(4)}, \label{eq-ord-gen-S-topo}
\end{equation}
where $S_{\alpha}^{(m)} = S_{\alpha}^{AB}$ in the four cases $(m)$
of partitioning the system shown in Fig. \ref{fig-3}. The
characteristic linear dimensions are the extension $D$ and the
thickness $d$ of the subsystem $A$ in all cases. To obtain a
meaningful topological measure, these dimensions need to be
macroscopic ($D > d \gg 1$).

The topological R\'{e}nyi entropy $S_{\alpha}^T $ is non-zero if and
only if the given state exhibits topological order \cite{TE,
Flammia}. For the TCM with external magnetic field, $S_{\alpha}^T =
0$ for the disordered ground state $| \Uparrow \, \rangle$ and
$S_{\alpha}^T = 2$ for the topologically ordered ground state $| 0
\rangle$. In this paper, we demonstrate that $S_{\alpha}^T $ detects
the presence of topological order in the entire topologically
ordered phase at $\lambda < \lambda_C$. Note though that
$S_{\alpha}^T$ is independent of $\alpha$ at the fixed point of a
generic non-chiral phase \cite{Flammia}. Since the fixed points of
different topologically ordered phases do not necessarily have
distinct values of $S_{\alpha}^T$, the topological R\'{e}nyi entropy
is unable to provide a complete characterization of a topologically
ordered phase.

\subsection{$\mathbb{Z}_2$ lattice gauge theory} \label{sec-ord-gau}

Since the TCM is perturbed with external fields $\hat{\sigma}_i^z$
that commute with the plaquettes $\hat{B}_p$ and the topological
operators $\hat{Z}_{1,2}$, the ground state $| \Omega (\lambda)
\rangle$ belongs to the lowest-energy sector with $Z_1 = Z_2 = +1$
and $B_p = +1$ ($\forall p$) for all values of $\lambda$. If we only
consider this sector, the gauge structure constraint $\hat{B}_p |
\Psi \rangle = | \Psi \rangle$ is enforced on all states $| \Psi
\rangle$, therefore the model is equivalent to a $\mathbb{Z}_2$
lattice gauge theory.

\begin{figure}[h!]
\centering
\includegraphics[width=6.4cm]{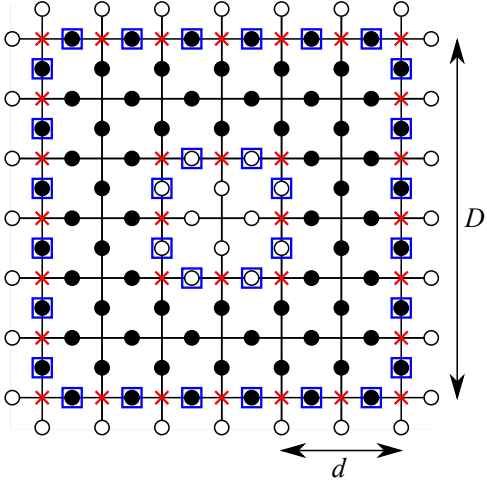}
\caption{(Color online) Illustration of the subsystems in case $(1)$
with dimensions $D = 6$ and $d = 2$. Spins are either in subsystem
$A$ (black circles) or in subsystem $B$ (white circles). Stars in
the set $C$ are marked by red crosses, and spins in the subsystem
$\partial A$ are marked by blue rectangles. The boundary contains $n
= 2$ closed loops on the real lattice with a combined length $L =
32$. \label{fig-4}}
\end{figure}

An arbitrary state $| \Psi \rangle$ within the gauge theory can be
expressed as a superposition of loop configurations. Each
configuration is a finite set of closed loops on the dual lattice,
and the spins on the loops are flipped with respect to the remaining
ones. These properties motivate us to introduce a modified
definition for the topological R\'{e}nyi entropy in which the
subsystem $A$ is substituted by the boundary $\partial A$ between
$A$ and $B$ in each case $(m)$ of partitioning the system. This
means that $S_{\alpha}^{(m)} = S_{\alpha}^{\partial A,
\overline{\partial A}} \equiv S_{\alpha}^{\partial A}$ in Eq.
(\ref{eq-ord-gen-S-topo}). Formally, we define $C$ as the set of
star operators acting on both subsystems $A$ and $B$, and $\partial
A$ as the set of spins that are only acted upon by stars in $C$. For
an illustration of this, see Fig. \ref{fig-4}. The boundary
$\partial A$ is always a finite set of closed loops on the real
lattice: the number of loops is $n = 2$ in the cases $(1)$ and
$(4)$, while it is $n = 1$ in the cases $(2)$ and $(3)$. Since a
loop on the real lattice and a loop on the dual lattice can only
intersect at an even number of points, there are an even number of
spins flipped on each loop of $\partial A$. This topological
constraint ensures that the modified $S_{\alpha}^T$ has similar
properties to the standard one. For example, it is still true that
$S_{\alpha}^T = 0$ for $| \Uparrow \, \rangle$ and $S_{\alpha}^T =
2$ for $| 0 \rangle$.

The calculations in the rest of the paper are immensely simplified
by using the modified definition for $S_{\alpha}^T$. Since the group
generated by the star operators acting exclusively on the boundary
subsystem $\partial A$ only contains the identity, the reduced
density matrix $\rho_{\partial A}$ is diagonal in the basis of the
physical spins $\sigma_i^z$. Each diagonal element $(\rho_{\partial
A})_{\Sigma \Sigma}$ gives the probability that $| \Psi \rangle$
realizes a given spin configuration $\{ \Sigma_i^z = \pm 1 \}$ in
$\partial A$. Equivalently, if we choose a random loop configuration
according to the probability distribution given by the state $| \Psi
\rangle$, the probability of the spin configuration $\{ \Sigma_i^z
\}$ in $\partial A$ is $P[ \{ \Sigma_i^z \} ] = (\rho_{\partial
A})_{\Sigma \Sigma}$. If we then choose two random loop
configurations according to the same distribution, the probability
of them having the same spin configuration in $\partial A$ is
\begin{equation}
\mathcal{P} = \sum_{\Sigma} P \left[ \{ \Sigma_i^z \} \right]^2 =
\sum_{\Sigma} \left( \rho_{\partial A} \right)_{\Sigma \Sigma}^2 =
\mathrm{Tr} \left[ \hat{\rho}_{\partial A}^2 \right].
\label{eq-ord-gau-prob}
\end{equation}
This result motivates us to consider the topological R\'{e}nyi
entropy of order $2$. In terms of the probabilities
$\mathcal{P}^{(m)}$ in the four cases $(m)$ of partitioning the
system, this quantity takes the form $S_2^T = \log_2
[\mathcal{P}^{(1)} \mathcal{P}^{(4)} / \mathcal{P}^{(2)}
\mathcal{P}^{(3)}]$.

We can now develop an intuitive understanding of the phase
transition by considering the two limiting cases. In the
topologically ordered ground state at $\lambda \ll 1$, the spin
loops are deconfined and all possible loop configurations are
equally probable. This means that the allowed spin configurations in
the subsystem $\partial A$ also share the same probability: the
inverse number of allowed spin configurations. It is important that
the number of boundary loops is $n = 2$ in the cases $(1)$ and
$(4)$, while it is $n = 1$ in the cases $(2)$ and $(3)$. The cases
$(1)$ and $(4)$ are therefore more constrained and have less allowed
spin configurations in $\partial A$. This implies
$\mathcal{P}^{(1)}, \mathcal{P}^{(4)} > \mathcal{P}^{(2)},
\mathcal{P}^{(3)}$ and $S_2^T > 0$. More precisely, since the
constraint on each boundary loop reduces the number of allowed spin
configurations by a factor of $2$, the topological R\'{e}nyi entropy
is given by $S_2^T = n^{(1)} - n^{(2)} - n^{(3)} + n^{(4)} = 2$. In
the disordered ground state at $\lambda \gg 1$, the spin loops are
confined and only the loop configurations with small spin loops have
significant probabilities. On the other hand, the small spin loops
in these loop configurations correspond to local disturbances
(nearby spin flips) in the spin configurations of the boundary
subsystem $\partial A$. This means that the probability $\mathcal
P^{(m)}$ in each case $(m)$ can be written as a product over the
small sections of the boundary loops, therefore $\log_2
\mathcal{P}^{(m)}$ is proportional to the length of the boundary.
Since the combined boundary length of the cases $(1)$ and $(4)$ is
equal to the combined boundary length of the cases $(2)$ and $(3)$,
the topological R\'{e}nyi entropy vanishes: $S_2^T = \log_2
[\mathcal{P}^{(1)} \mathcal{P}^{(4)} / \mathcal{P}^{(2)}
\mathcal{P}^{(3)}] = 0$.

\subsection{Formula for the R\'{e}nyi entropy} \label{sec-ord-ent}

Now we capitalize on the simplifications described above, and derive
the R\'{e}nyi entropy $S_2^{\partial A}$ for an arbitrary state $|
\Psi \rangle$ within the gauge theory. In the most general case,
$\partial A$ consists of $n$ closed loops on the real lattice, and
the loops have a combined length $L$. This means that they contain
$L$ spins and $L$ stars acting on these spins (see Fig.
\ref{fig-4}). Since there is a constraint on each loop due to the
gauge structure, only $L - n$ spins are independent. If we label
these spins with $1 \leq i \leq L - n$, the $2^{L-n}$ non-zero
diagonal elements of $\rho_{\partial A}$ give the probabilities of
$| \Psi \rangle$ realizing the $2^{L-n}$ respective spin
configurations $\{ \Sigma_i^z \}$. Since the projection operator
onto the spin configuration $\{ \Sigma^z_i \}$ is given by $2^{n-L}
\prod_{i} (1 + \Sigma_i^z \hat{\sigma}_i^z) $, the corresponding
diagonal element reads
\begin{equation}
\left( \rho_{\partial A} \right)_{\Sigma \Sigma} = \frac{1}
{2^{L-n}} \langle \Psi | \left[ \prod_{i=1}^{L-n} (1 + \Sigma_i^z
\hat{\sigma}_i^z) \right] | \Psi \rangle. \label{eq-ord-ent-diag}
\end{equation}
When expanding the product in Eq. (\ref{eq-ord-ent-diag}) and
summing the squares of the resulting expressions for
$(\rho_{\partial A})_{\Sigma \Sigma}$, the cross-terms cancel each
other, and we obtain
\begin{equation}
\mathrm{Tr} \left[ \hat{\rho}_{\partial A}^2 \right] = \frac{1}
{2^{L-n}} \sum_{\{ q_i = 0,1 \}} \langle \Psi | \left[
\prod_{i=1}^{L-n} \left( \hat{\sigma}_i^z \right)^{q_i} \right] |
\Psi \rangle^2, \label{eq-ord-ent-tr}
\end{equation}
where the sum is over all the $2^{L-n}$ configurations $\{ q_i = 0,1
\}$, and hence over all possible products of the $L - n$ independent
spin operators $\hat{\sigma}_i^z$. If the edge occupied by the spin
$i$ connects the stars $s_{i,1}$ and $s_{i,2}$, the corresponding
spin operator becomes $\hat{\sigma}_i^z = \hat{A}_{s_{i,1}}^x
\hat{A}_{s_{i,2}}^x$. In terms of the quasi-spin operators
$\hat{A}_s^x$, the R\'{e}nyi entropy then takes the form
\begin{equation}
S_2^{\partial A} = (L - n) - \log_2 \sum_{\{ q_i = 0,1 \}} \langle
\Psi | \left[ \prod_{i=1}^{L-n} \left( \hat{A}_{s_{i,1}}^x
\hat{A}_{s_{i,2}}^x \right)^{q_i} \right] | \Psi \rangle^2.
\label{eq-ord-ent-S-1}
\end{equation}
This expression has an entirely precise notation, but it is
cumbersome to use for calculating $S_2^{\partial A}$. To derive a
more intuitive expression with a less precise notation, we expand
the sum in Eq. (\ref{eq-ord-ent-S-1}) around the trivial
configuration $\{ q_i = 0 \}$. Exploiting $(\hat{A}_s^x)^2 = 1$, the
R\'{e}nyi entropy then becomes
\begin{eqnarray}
S_2^{\partial A} &=& (L - n) - \log_2 \Bigg{[} 1 + \sum_{s_1, s_2}
\langle \Psi | \hat{A}_{s_1}^x \hat{A}_{s_2}^x | \Psi \rangle^2
\nonumber \\
&+& \sum_{s_1, s_2, s_3, s_4} \langle \Psi | \hat{A}_{s_1}^x
\hat{A}_{s_2}^x \hat{A}_{s_3}^x \hat{A}_{s_4}^x | \Psi \rangle^2\ +
\ldots \Bigg{]}, \qquad \label{eq-ord-ent-S-2}
\end{eqnarray}
where the sum inside the logarithm contains all $2^{L-n}$ possible
products with an even number of quasi-spin operators $\hat{A}_s^x$
chosen from each closed loop of the subsystem $\partial A$.

To understand how Eq. (\ref{eq-ord-ent-S-2}) works, we consider the
two limiting ground states $| \Uparrow \, \rangle$ and $| 0
\rangle$. In the first case, we have $\langle \Uparrow \, |
\hat{A}_{s_1}^x \hat{A}_{s_2}^x \ldots \hat{A}_{s_{2r}}^x | \Uparrow
\, \rangle = \langle \Uparrow \, | \hat{\sigma}_{i_1}^z
\hat{\sigma}_{i_2}^z \ldots \hat{\sigma}_{i_q}^z | \Uparrow \,
\rangle = 1$ for all expectation values because $| \Uparrow \,
\rangle$ has $\sigma_i^z = +1$ for all $i$. The sum inside the
logarithm becomes $2^{L-n}$, and the R\'{e}nyi entropy
$S_2^{\partial A}$ vanishes, as expected for a product state. In the
second case, $| 0 \rangle$ has $A_s^z = +1$ for all $s$, therefore
$\langle 0 | \hat{A}_{s_1}^x \hat{A}_{s_2}^x \ldots
\hat{A}_{s_{2r}}^x | 0 \rangle = 0$ for all expectation values. The
only exception is the trivial one: $\langle 0 | 0 \rangle = 1$. The
sum inside the logarithm is $1$, and the R\'{e}nyi entropy is
$S_2^{\partial A} = L - n$. When extracting the topological
contribution, the terms $\propto L$ cancel because $L^{(1)} +
L^{(4)} = L^{(2)} + L^{(3)}$ (see Fig. \ref{fig-3}). On the other
hand, the cases $(1)$ and $(4)$ have $n = 2$, while the cases $(2)$
and $(3)$ have $n = 1$, therefore the topological R\'{e}nyi entropy
is finite: $S_2^T = 2$.

\section{Phase transition in the quasi-1D case} \label{sec-1D}

In this section, we set $\kappa \ll 1$ in Eq. (\ref{eq-gen-H-old}):
this means that the spins on the vertical edges experience much
smaller magnetic fields than those on the horizontal edges. The
magnetic fields on the vertical edges only become important in the
$\lambda \gg 1$ limit, therefore the phase transition between
topological order and disorder at $\lambda \sim 1$ occurs due to a
competition between the star operators $\hat{A}_s^z$ and the
magnetic fields on the horizontal edges. When investigating this
phase transition, the magnetic fields on the vertical edges can be
neglected ($\kappa \rightarrow 0$), and the quasi-spin Hamiltonian
in Eq. (\ref{eq-gen-H-new}) becomes
\begin{equation}
\hat{H} = - \sum_s \hat{A}_s^z - \lambda \sum_{\langle s,s' \rangle
\in h} \hat{A}_s^x \hat{A}_{s'}^x. \label{eq-1D-H}
\end{equation}
Since there are only Ising couplings on the horizontal edges, this
Hamiltonian is the direct sum of $N$ independent 1D TFIM copies
along the horizontal chains of the lattice \cite{Yu}. The system is
therefore exactly solvable for all values of $\lambda$.

If we consider any of the independent 1D horizontal chains and label
the $N$ stars on the chain with $1 \leq l \leq N$, the Hamiltonian
of the corresponding 1D TFIM reads
\begin{equation}
\hat{H} = - \sum_{l=1}^{N} \left( \hat{A}_l^z + \lambda \hat{A}_l^x
\hat{A}_{l-1}^x \right), \label{eq-1D-H-old}
\end{equation}
where the periodic boundary conditions are taken into account by
$A_0 \equiv A_N$. This Hamiltonian can be solved by following a
standard procedure \cite{Barouch-1}. We first map the quasi-spins
$A_l$ to fermions via the Jordan-Wigner transformation
\begin{equation}
\hat{A}_l^z = 1 - 2 c_l^{\dag} c_l, \quad \hat{A}_l^- = \left(
\hat{A}_l^+ \right)^{\dag} = c_l^{\dag} e^{i \pi \sum_{j=1}^{l-1}
c_j^{\dag} c_j}, \label{eq-1D-JW}
\end{equation}
where $c_l^{\dag}$ and $c_l$ are standard fermionic creation and
annihilation operators. The translational symmetry is then exploited
by the Fourier transform
\begin{equation}
c_l = \frac{1} {\sqrt{N}} \sum_k e^{ikl} c_k, \label{eq-1D-FT}
\end{equation}
where the sum is over the momenta $k_m = \pi (2m - 1) / N$ with $1
\leq m \leq N$. We finally introduce new fermionic operators via the
Bogoliubov transformation
\begin{equation}
c_k = \cos \theta_k \gamma_k + i \sin \theta_k \gamma_{-k}^{\dag},
\label{eq-1D-bog}
\end{equation}
and the Hamiltonian in Eq. (\ref{eq-1D-H-old}) becomes
\begin{equation}
\hat{H} = \sum_k \Lambda_{k} \left( 2 \gamma_k^{\dag} \gamma_k - 1
\right), \label{eq-1D-H-new}
\end{equation}
where $\gamma_k^{\dag}$ and $\gamma_k$ correspond to independent
fermionic quasi-particles. The energies of these quasi-particles are
proportional to $\Lambda_k = \sqrt{\epsilon_k^2 + \lambda^2 \sin^2
k}$ with $\epsilon_k \equiv 1 - \lambda \cos k$, and the mixing
angle appearing in the Bogoliubov transformation is $\theta_k =
\tan^{-1} [\lambda \sin k / (\epsilon_k + \Lambda_k)]$.

The ground state $| \Omega (\lambda) \rangle$ of the Hamiltonian in
Eq. (\ref{eq-1D-H}) is the direct product of $N$ independent copies
of the 1D ground state $| \Omega_0 \rangle$. The 1D ground state is
defined by $\gamma_k | \Omega_0 \rangle = 0$ for all $k$, therefore
its two-operator expectation values in the position representation
are given by
\begin{eqnarray}
\langle c_l^{\dag} c_{l'} \rangle &\equiv& \langle \Omega_0 |
c_l^{\dag} c_{l'} | \Omega_0 \rangle = \frac{1}{N} \sum_{k, k'}
e^{-ikl +ik'l'} \langle \Omega_0 | c_k^{\dag} c_{k'} | \Omega_0
\rangle \nonumber \\
&=& \frac{1}{N} \sum_k e^{-ik (l - l')} \sin^2 \theta_k,
\nonumber \\
\langle c_l c_{l'}^{\dag} \rangle &=& \frac{1}{N} \sum_k e^{-ik (l -
l')} \cos^2 \theta_k, \label{eq-1D-cc} \\
\langle c_l c_{l'} \rangle &=& \frac{i}{N} \sum_k e^{-ik (l - l')}
\sin \theta_k \cos \theta_k, \nonumber \\
\langle c_l^{\dag} c_{l'}^{\dag} \rangle &=& - \frac{i}{N} \sum_k
e^{-ik (l - l')} \sin \theta_k \cos \theta_k. \nonumber
\end{eqnarray}
To calculate the R\'{e}nyi entropy, we need to evaluate the
quasi-spin expectation values appearing in Eq.
(\ref{eq-ord-ent-S-2}). These expectation values are products of
independent 1D expectation values $\langle \Omega_0 |
\hat{A}_{l_1}^x \hat{A}_{l_2}^x \ldots \hat{A}_{l_{2r-1}}^x
\hat{A}_{l_{2r}}^x | \Omega_0 \rangle$, where each pair of
$\hat{A}_l^x$ operators can be expressed in terms of the fermionic
operators as
\begin{equation}
\hat{A}_l^x \hat{A}_{l'}^x = \left( c_l^{\dag} + c_l \right)
\prod_{j=l}^{l'-1} \left( 1 - 2 c_j^{\dag} c_j \right) \left(
c_{l'}^{\dag} + c_{l'} \right). \label{eq-1D-S}
\end{equation}
Similarly, the quasi-spin expectation value in the Wilson loop for a
$D \times D$ square region $R$ becomes
\begin{equation}
W_R = \langle \Omega_0 | \prod_{l=1}^{D} \hat{A}_l^z | \Omega_0
\rangle^D = \langle \Omega_0 | \prod_{l=1}^{D} \left( 1 - 2
c_l^{\dag} c_l \right) | \Omega_0 \rangle^D. \label{eq-1D-W}
\end{equation}
Using the identity $1 - 2 c_l^{\dag} c_l = (c_l^{\dag} + c_l)
(c_l^{\dag} - c_l)$, the quasi-spin operator products appearing in
both the R\'{e}nyi entropy and the Wilson loop can then be written
as simple products of $c_l^{\dag} \pm c_l$ operators. On the other
hand, the expectation values of these products can be reduced to the
two-operator expectation values given in Eq. (\ref{eq-1D-cc}) by
using Wick's theorem \cite{Barouch-2}.

\begin{figure}[t!]
\centering
\includegraphics[width=8.6cm]{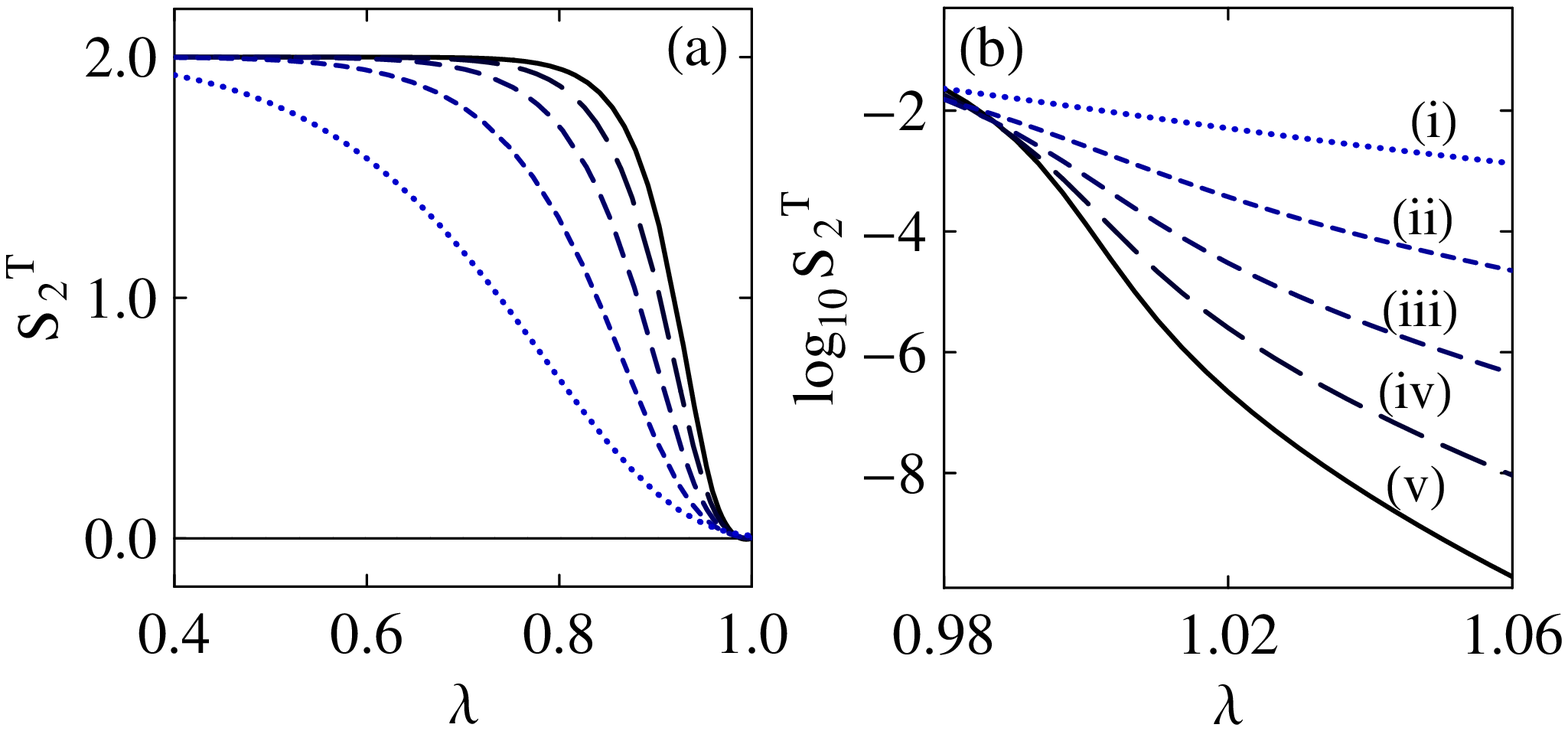}
\caption{(Color online) Topological R\'{e}nyi entropy $S_2^T$ as a
function of magnetic field $\lambda$ below (a) and above (b) the
critical point at $\lambda_C = 1$. Five system sizes are plotted
with $d = D/3$ for each: $N = 40$ and $D = 6$ (i); $N = 80$ and $D =
12$ (ii); $N = 120$ and $D = 18$ (iii); $N = 160$ and $D = 24$ (iv);
$N = 200$ and $D = 30$ (v). \label{fig-5}}
\end{figure}

The exact dependence of the topological R\'{e}nyi entropy on the
magnetic field is plotted in Fig. \ref{fig-5}. There are two phases:
a topologically ordered phase at small $\lambda$ and a disordered
phase at large $\lambda$. These phases are separated by a clear
phase transition at $\lambda = \lambda_C = 1$, which coincides with
the well-known critical point of the 1D TFIM \cite{Barouch-1}. If we
gradually increase $\lambda$, the topological R\'{e}nyi entropy
drops to zero around $\lambda_C$. This transition becomes sharper if
we increase the system size $N$ as well as the dimensions $D$ and
$d$ of the subsystems, therefore we argue that $S_2^T$ is
discontinuous in the thermodynamic limit. The topological R\'{e}nyi
entropy is then constant in both limiting phases: the topologically
ordered phase at $\lambda < \lambda_C$ is characterized by $S_2^T =
2$, while the disordered phase at $\lambda > \lambda_C$ is
characterized by $S_2^T = 0$.

\begin{figure}[t!]
\centering
\includegraphics[width=8.4cm]{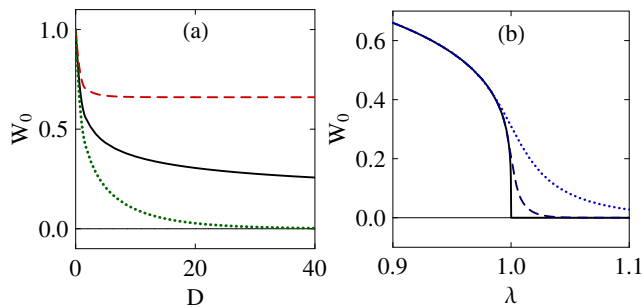}
\caption{(Color online) (a) Reduced Wilson loop $W_0$ as a function
of region size $D$ for $\lambda = 0.9 < \lambda_C$ (red dashed
line), $\lambda = 1.0 = \lambda_C$ (black solid line), and $\lambda
= 1.1 > \lambda_C$ (green dotted line). The system size is $N =
500$. (b) Reduced Wilson loop $W_0$ as a function of magnetic field
$\lambda$ for system sizes $N = 5D = 100$ (dotted line) and $N = 5D
= 500$ (dashed line). The solid line is the result in Eq.
(\ref{eq-dis-per-W-2}) for the thermodynamic limit. \label{fig-6}}
\end{figure}

The analogous exact behavior of the Wilson loop is illustrated in
Fig. \ref{fig-6}. In the topologically ordered phase at $\lambda <
\lambda_C$, the reduced Wilson loop $W_0 \equiv W_R^{1/D}$
approaches a finite constant in the $D \rightarrow \infty$ limit.
This implies $W_R \propto \exp (- \beta D)$ and the presence of
topological order. In the disordered phase at $\lambda > \lambda_C$,
$W_0$ decays exponentially with $D$. This implies $W_R \propto \exp
(- \beta D^2)$ and the absence of topological order. By looking at
the dependence $W_0 (\lambda)$ for a sufficiently large value of
$D$, we can establish that the critical point separating the two
different behaviors is indeed at $\lambda_C = 1$. The results
obtained for the topological R\'{e}nyi entropy and the Wilson loop
are therefore consistent with each other.

\section{Phase transition in the actual 2D case} \label{sec-2D}

In this section, we set $\kappa = 1$ in Eq. (\ref{eq-gen-H-old}):
this means that the spins on the horizontal and the vertical edges
experience the same magnetic field. Up to an irrelevant additive
constant, the quasi-spin Hamiltonian in Eq. (\ref{eq-gen-H-new})
becomes
\begin{equation}
\hat{H} = \sum_s \left( 1 - \hat{A}_s^z \right) - \lambda
\sum_{\langle s,s' \rangle} \hat{A}_s^x \hat{A}_{s'}^x,
\label{eq-2D-H}
\end{equation}
and the system is equivalent to the standard 2D TFIM. Since the
Hamiltonian in Eq. (\ref{eq-2D-H}) is not exactly solvable in
general, we use perturbation theories around the exactly solvable
limits at $\lambda = 0$ and $\lambda \rightarrow \infty$. The
corresponding calculations are most efficiently performed by the
method of perturbative continuous unitary transformations (PCUT).
The general method is discussed in the literature \cite{Dusuel,
PCUT} and we illustrate its use by the example of our particular
problem.

\subsection{Perturbation theory at small magnetic field}
\label{sec-2D-sm}

In the limit of $\lambda \ll 1$, it is useful to work in the
quasi-spin representation because the unperturbed ground state $| 0
\rangle$ is then a product state. The perturbation theory is based
on Eq. (\ref{eq-2D-H}), where the second term is treated as a
perturbation in the small parameter $\lambda \ll 1$. Using the PCUT
procedure described in Appendix \ref{sec-app-sm}, we obtain
corrections to the R\'{e}nyi entropy $S_2^{\partial A}$ for each
case of partitioning in Fig. \ref{fig-3} and the Wilson loop $W_R$
for a square region $R$. The R\'{e}nyi entropy after the first three
corrections reads
\begin{eqnarray}
S_2^{\partial A} &=& (L - n) - \frac{L}{\ln 2} \left[
\frac{\lambda^2}{4} + \frac{63\lambda^4}{64} +
\frac{503\lambda^6}{96} + O(\lambda^8) \right] \nonumber \\
&-& \frac{K}{\ln 2} \left[ \frac{27\lambda^4}{64} +
\frac{737\lambda^6}{256} + O(\lambda^8) \right], \label{eq-2D-sm-S}
\end{eqnarray}
where the boundary $\partial A$ contains $n$ closed loops with a
combined length $L$ and a total number of $K$ corners that are
sufficiently far away from each other. The analogous expression for
the Wilson loop after the first three corrections is
\begin{eqnarray}
W_R &=& \exp \bigg{\{} - L \left[ \frac{\lambda^2}{8} +
\frac{\lambda^4}{2} + \frac{7697\lambda^6}{3072} + O(\lambda^8)
\right] \nonumber \\
&+& K \left[ \frac{3\lambda^4}{32} + \frac{89\lambda^6}{128} +
O(\lambda^8) \right] \bigg{\}}, \label{eq-2D-sm-W}
\end{eqnarray}
where the square region $R$ has a boundary length $L = 4D$ and a
corner number $K = 4$.

Since $K$ is merely an $O(1)$ constant, the corrections inside the
exponential of Eq. (\ref{eq-2D-sm-W}) are linearly proportional to
the region dimension $D$. The Wilson loop has therefore a functional
form $W_R \propto \exp (-\beta D)$ that shows the presence of
topological order. Since the corrections to the R\'{e}nyi entropy
are all linearly proportional to either $L$ or $K$, the corrections
to the topological contribution $\propto n$ vanish. When calculating
the topological R\'{e}nyi entropy, the corrections $\propto L, K$
cancel because the combined values of $L$ and $K$ in the cases $(1)$
and $(4)$ match those in the cases $(2)$ and $(3)$ (see Fig.
\ref{fig-3}). The topological R\'{e}nyi entropy is therefore
constant up to the third correction: $S_2^T = 2 + O(\lambda^8)$ in
the $\lambda \ll 1$ phase.

\subsection{Perturbation theory at large magnetic field}
\label{sec-2D-lg}

In the limit of $\lambda \gg 1$, it is useful to return to the
physical spin representation because the unperturbed ground state $|
\Uparrow \, \rangle$ is then a product state. Up to an irrelevant
additive constant and an overall multiplicative factor
$\lambda^{-1}$, the 2D TFIM Hamiltonian in Eq. (\ref{eq-2D-H})
becomes
\begin{equation}
\hat{H} = \sum_i \left( 1 - \hat{\sigma}_i^z \right) - \lambda^{-1}
\sum_s \prod_{i \in s} \hat{\sigma}_i^x. \label{eq-2D-lg-H}
\end{equation}
The perturbation theory is based on Eq. (\ref{eq-2D-lg-H}), where
the second term is treated as a perturbation in the small parameter
$\lambda^{-1} \ll 1$. Using the PCUT procedure described in Appendix
\ref{sec-app-lg}, the R\'{e}nyi entropy after the first three
corrections is
\begin{eqnarray}
S_2^{\partial A} &=& \frac{L}{\ln 2} \left[ \frac{\lambda^{-2}}{32}
+ \frac{\lambda^{-4}}{1024} +
\frac{115\lambda^{-6}}{2359296} + O(\lambda^{-8}) \right] \nonumber \\
&-& \frac{K}{\ln 2} \left[ \frac{35\lambda^{-6}}{4718592} +
O(\lambda^{-8}) \right]. \label{eq-2D-lg-S}
\end{eqnarray}
Since the corrections to the R\'{e}nyi entropy are all linearly
proportional to either $L$ or $K$, the corrections to the
topological contribution $\propto n$ vanish. The topological
R\'{e}nyi entropy is therefore constant zero up to the third
correction: $S_2^T = O(\lambda^{-8})$ in the $\lambda \gg 1$ phase.

To obtain a non-zero result for the Wilson loop expectation value
$\langle \Omega (\lambda) | \hat{W}_R | \Omega (\lambda) \rangle$,
we need to consider higher orders of perturbation theory. Since
$\hat{W}_R$ is a product of $D^2$ star operators $\hat{A}_s^z$, the
first non-zero contribution to $W_R$ appears at order $D^2 / 2$ in
perturbation theory. At this order, $\hat{W}_R$ links order $D^2 /
2$ states to each other, therefore $W_R \propto \lambda^{-D^2}$.
This result can be rearranged into the form $W_R \propto \exp [- \ln
(\lambda) \, D^2]$ that shows the absence of topological order.

\subsection{Discussion of the phase transition}
\label{sec-2D-dis}

The results of the perturbation theories indicate two distinct
phases around the limits $\lambda = 0$ and $\lambda \rightarrow
\infty$. The phase at $\lambda \ll 1$ is topologically ordered
because the topological R\'{e}nyi entropy is non-zero and the Wilson
loop follows a perimeter law: $W_R \propto \exp (- \beta D)$.
Conversely, the phase at $\lambda \gg 1$ is disordered because the
topological R\'{e}nyi entropy is zero and the Wilson loop follows an
area law: $W_R \propto \exp (- \beta D^2)$.

The topological distinctness implies at least one phase transition
between the two limiting phases, and we argue that there can only be
one phase transition. Recall that the TCM with external field is
equivalent to the standard 2D TFIM when $\kappa = 1$. In particular,
the quantities $S_2^T$ and $W_R$ that describe topological order can
be expressed in terms of the 2D TFIM correlation functions. A phase
transition is therefore only possible at the critical point of the
2D TFIM, which has been determined by various numerical methods
\cite{Ising} to be at $\lambda_C \approx 0.33$. This critical field
is also consistent with previous numerical studies on the TCM with
external field \cite{Hamma-3, Trebst}.

It is clear that the perturbation theories around the two limits
need to break down at $\lambda = \lambda_C$. On the other hand, the
results of the perturbation theories hold because the expansions in
Eqs. (\ref{eq-2D-sm-S}), (\ref{eq-2D-sm-W}), and (\ref{eq-2D-lg-S})
have particular structures: they each contain two power series in
$\lambda$ that are proportional to the boundary length $L$ and the
corner number $K$. It is plausible that higher order corrections
preserve this form and only add further terms to the respective
power series. Terms that are not linearly proportional to either $L$
or $K$ only appear when the order of the perturbation theory exceeds
the dimensions $D$ and $d$ of the subsystems (regions). Since these
dimensions are macroscopic in the thermodynamic limit, the
perturbation theories can only break down at infinitely large
orders. These in turn become important at the radii of convergence
where the series actually diverge. If we write the power series in
Eqs. (\ref{eq-2D-sm-S}) and (\ref{eq-2D-sm-W}) as
$\sum_{k=1}^{\infty} a_k \lambda^{2k}$ and those in Eq.
(\ref{eq-2D-lg-S}) as $\sum_{k=1}^{\infty} b_k \lambda^{-2k}$, the
critical field $\lambda_C$ is given by
\begin{equation}
\lambda_C = \lim_{k \rightarrow \infty} \sqrt{ \left|
\frac{a_k}{a_{k+1}} \right| } = \lim_{k \rightarrow \infty} \sqrt{
\left| \frac{b_{k+1}}{b_k} \right| } \, . \label{eq-2D-dis-crit}
\end{equation}
Although it is not possible to determine these limits from a
finite-order perturbation theory, we can give estimates for the
critical field by looking at the first couple of terms and
calculating analogous quantities. The resulting estimates are
summarized in Table \ref{tab-1}: they suggest $0.2 \lesssim
\lambda_C \lesssim 0.5$. This range is fully consistent with
$\lambda_C \approx 0.33$.

\begin{table}[h]
\begin{tabular*}{0.45\textwidth}{@{\extracolsep{\fill}}c c c c}
\hline \hline
\multicolumn{2}{c}{Estimates for $\lambda_C$}                      &   $\sqrt{a_1 / a_2}$   &   $\sqrt{a_2 / a_3}$   \\
\hline
\multirow{2}{*}{Eq. (\ref{eq-2D-sm-S})}   &   Series $\propto L$   &   0.504                &   0.433                \\
                                          &   Series $\propto K$   &   -                    &   0.383                \\
\hline
\multirow{2}{*}{Eq. (\ref{eq-2D-sm-W})}   &   Series $\propto L$   &   0.500                &   0.447                \\
                                          &   Series $\propto K$   &   -                    &   0.367                \\
\hline  \hline
\\
\hline  \hline
\multicolumn{2}{c}{Estimates for $\lambda_C$}                      &   $\sqrt{b_2 / b_1}$   &   $\sqrt{b_3 / b_2}$   \\
\hline
\multirow{2}{*}{Eq. (\ref{eq-2D-lg-S})}   &   Series $\propto L$   &   0.177                &   0.223                \\
                                          &   Series $\propto K$   &   -                    &   -                    \\
\hline  \hline
\end{tabular*}
\caption{Estimates for the critical field $\lambda_C$ obtained from
the power series of Eqs. (\ref{eq-2D-sm-S}), (\ref{eq-2D-sm-W}), and
(\ref{eq-2D-lg-S}). \label{tab-1}}
\end{table}

The most remarkable result of this section is that the topological
R\'{e}nyi entropy is constant in both limiting phases: $S_2^T = 2$
in the topologically ordered phase and $S_2^T = 0$ in the disordered
phase. This happens because the perturbative corrections to
$S_2^{\partial A}$ do not contain any topological contributions
$\propto n$ in Eqs. (\ref{eq-2D-sm-S}) and (\ref{eq-2D-lg-S}). The
topological R\'{e}nyi entropy is therefore an exclusive function of
the phase: it can only change if a phase transition takes place. We
argue that $S_2^T$ is a good probe of topological order with the
potential to characterize topologically ordered phases.

\section{Comparison and discussion} \label{sec-dis}

\subsection{Perturbation theories in the quasi-1D case} \label{sec-dis-per}

Although the quasi-1D case is exactly solvable, it is instructive to
treat it with perturbation theories as well: the results obtained
this way are directly comparable with those in the actual 2D case.
Using a modified version of the PCUT procedures described in the
Appendix, we find analogous expressions to those in Eqs.
(\ref{eq-2D-sm-S}), (\ref{eq-2D-sm-W}), and (\ref{eq-2D-lg-S}).
Without including the detailed calculations, the R\'{e}nyi entropies
after the first three corrections in the two limiting regimes are
\begin{eqnarray}
S_2^{\partial A} &=& (L - n) - \frac{L'}{\ln 2} \left[
\frac{\lambda^2}{4} + \frac{7\lambda^4}{64} + \frac{5\lambda^6}{96}
+ O(\lambda^8) \right] \nonumber \\
&+& \frac{H'}{\ln 2} \left[ \frac{5\lambda^4}{64} +
\frac{3\lambda^6}{32} + O(\lambda^8) \right] \quad (\lambda \ll 1),
\label{eq-dis-per-S-sm}
\end{eqnarray}
\begin{eqnarray}
S_2^{\partial A} &=& \frac{L'}{\ln 2} \left[ \frac{\lambda^{-2}}{8}
+ \frac{\lambda^{-4}}{32} + \frac{47\lambda^{-6}}{3072} +
O(\lambda^{-8}) \right] \nonumber \\
&+& \frac{H'}{\ln 2} \left[ \frac{\lambda^{-2}}{8} +
\frac{7\lambda^{-4}}{128} +\frac{107\lambda^{-6}}{3072} +
O(\lambda^{-8}) \right] \nonumber \\
&& (\lambda \gg 1), \label{eq-dis-per-S-lg}
\end{eqnarray}
where $L'$ is the combined horizontal length of the boundary
$\partial A$, and $H'$ is the number of horizontal sections with a
non-zero length contributing to $L'$. Since the corrections to the
R\'{e}nyi entropies are all linearly proportional to either $L'$ or
$H'$, there are no topological corrections $\propto n$. The
topological R\'{e}nyi entropy is therefore constant $S_2^T = 2 +
O(\lambda^8)$ at $\lambda \ll 1$ and constant $S_2^T =
O(\lambda^{-8})$ at $\lambda \gg 1$.

When taking into account the first three corrections, the reduced
Wilson loop in the $\lambda \ll 1$ regime becomes
\begin{equation}
W_0 = \exp \bigg{\{} - \left[ \frac{\lambda^2}{4} +
\frac{\lambda^4}{8} + \frac{\lambda^6}{12} + O(\lambda^8) \right]
\bigg{\}}, \label{eq-dis-per-W-1}
\end{equation}
which indicates $W_R \propto \exp (-\beta D)$ and the presence of
topological order. In the $\lambda \gg 1$ regime, the first non-zero
contribution to $W_0$ appears at order $D/2$ in perturbation theory.
This contribution is $W_0 \propto \lambda^{-D} = \exp [- \ln
(\lambda) \, D]$, which indicates $W_R \propto \exp (-\beta D^2)$
and the absence of topological order. Note that the power series
inside the exponential of Eq. (\ref{eq-dis-per-W-1}) suggests that
$W_0$ takes the exact form
\begin{equation}
W_0 = \exp \left( - \sum_{k=1}^{\infty} \frac{\lambda^{2k}}{4k}
\right) = \left( 1 - \lambda^2 \right)^{1/4} \label{eq-dis-per-W-2}
\end{equation}
in the thermodynamic limit. This result is consistent with the
critical field $\lambda_C = 1$ obtained from the exact treatment.

\begin{table}[h]
\begin{tabular*}{0.45\textwidth}{@{\extracolsep{\fill}}c c c c}
\hline \hline
\multicolumn{2}{c}{Estimates for $\lambda_C$}                            &   $\sqrt{a_1 / a_2}$   &   $\sqrt{a_2 / a_3}$   \\
\hline
\multirow{2}{*}{Eq. (\ref{eq-dis-per-S-sm})}   &   Series $\propto L'$   &   1.512                &   1.449                \\
                                               &   Series $\propto H'$   &   -                    &   0.913                \\
\hline
Eq. (\ref{eq-dis-per-W-1})                     &   Series $\propto 1$    &   1.414                &   1.225                \\
\hline  \hline
\\
\hline  \hline
\multicolumn{2}{c}{Estimates for $\lambda_C$}                            &   $\sqrt{b_2 / b_1}$   &   $\sqrt{b_3 / b_2}$   \\
\hline
\multirow{2}{*}{Eq. (\ref{eq-dis-per-S-lg})}   &   Series $\propto L'$   &   0.500                &   0.700                \\
                                               &   Series $\propto H'$   &   0.661                &   0.798                \\
\hline  \hline
\end{tabular*}
\caption{Estimates for the critical field $\lambda_C$ obtained from
the power series of Eqs. (\ref{eq-dis-per-S-sm}),
(\ref{eq-dis-per-S-lg}), and (\ref{eq-dis-per-W-1}). \label{tab-2}}
\end{table}

The perturbative expansions in Eqs. (\ref{eq-dis-per-S-sm}),
(\ref{eq-dis-per-S-lg}), and (\ref{eq-dis-per-W-1}) each contain at
least one power series in $\lambda$. The critical field $\lambda_C$
marks the breakdown of the perturbation theories, and it is again
related to the appropriate radii of convergence. The estimates
obtained with the method of Sec. \ref{sec-2D-dis} are summarized in
Table \ref{tab-2}: they suggest $0.5 \lesssim \lambda_C \lesssim
1.5$. This range is fully consistent with $\lambda_C = 1$.

\subsection{Comparison with the actual 2D case} \label{sec-dis-com}

When discussing the phase transition in the actual 2D case, we
argued that it occurs at the critical point $\lambda_C \approx 0.33$
of the equivalent 2D TFIM and that the two limiting phases are
characterized by different constant values of the topological
R\'{e}nyi entropy. The argument only referred to the perturbation
theories and the equivalence with the 2D TFIM. On the other hand,
the quasi-1D case is more versatile because an exact solution is
available. The exact treatment of the quasi-1D case suggests a
behavior that is entirely analogous to our claims for the actual 2D
case: the phase transition occurs at the critical point $\lambda_C =
1$ of the equivalent 1D TFIM, and the topological R\'{e}nyi entropy
is constant in the two limiting phases.

A direct comparison between the respective perturbation theories
also provides evidence that the 1D and the 2D systems are similar in
terms of their phase transitions. The behaviors of the $\lambda_C$
estimates and their relations to the actual $\lambda_C$ are entirely
analogous in the two cases. First, the estimates are all reasonably
close to the actual $\lambda_C$. Second, the estimates converge
towards $\lambda_C$ as the order is increased. Third, the estimates
from the $\lambda \ll 1$ series generally overestimate, while those
from the $\lambda \gg 1$ series underestimate $\lambda_C$. These
similarities suggest that the phase transitions in the 1D and the 2D
cases are analogous, therefore the conclusions drawn from the exact
treatment in the quasi-1D case are applicable to the physically more
interesting actual 2D case as well.

\section{Conclusions} \label{sec-con}

In this paper, we investigated the quantum phase transition between
the topologically ordered and the disordered phases of the TCM with
external magnetic field. The variation in topological order was
probed via $S_2^T$: the topological R\'{e}nyi entropy of order $2$.
We determined the exact field dependence of $S_2^T$ in the
computationally simpler case (quasi-1D case) and established
perturbation theories in the physically more interesting case
(actual 2D case). It was demonstrated that $S_2^T$ takes distinct
values in the two phases and has a discontinuity at the quantum
phase transition. We therefore argue that $S_2^T$ is a good probe of
topological order that can effectively characterize topologically
ordered phases.

The equivalence between the quasi-1D case of our problem and the
exactly solvable 1D TFIM is a quite remarkable tool for obtaining
exact results. So far it has provided us with an exact treatment of
the quasi-1D case and a corresponding exact $S_2^T(\lambda)$
dependence. In perspective, such an exact treatment also makes it
possible to search for critical exponents that can reveal the
topological character of the quantum phase transition. Moreover, the
exact time dependence of the system far away from equilibrium can be
studied, as for example, in the case of a quantum quench
\cite{Tsomokos, Halasz}.

It is important to point out that the Hamiltonian in Eq.
(\ref{eq-gen-H-old}) preserves the $\mathbb{Z}_2$ gauge structure of
the bare TCM for all values of the magnetic field $\lambda$. This
gauge structure justifies the simplifying step of substituting the
subsystem $A$ by its boundary $\partial A$ when calculating $S_2^T$
(thin subsystem). Indeed, as long as the gauge structure is
preserved by the perturbation, the ground state can be expressed as
a superposition of loop configurations. For such a system, all the
relevant topological constraints are necessarily connected to the
subsystem boundary $\partial A$. For example, in our $\mathbb{Z}_2$
gauge theory, the topological constraint manifests itself in the
fact that there are an even number of spins flipped on each boundary
loop of $\partial A$. On the other hand, considering only the
boundary is the essential simplification we need for deriving the
R\'{e}nyi entropy formula in Eq. (\ref{eq-ord-ent-S-2}), which in
turn makes the exact treatment in the quasi-1D case and the
perturbation theories in the actual 2D case possible. The gauge
structure also explains why the topological R\'{e}nyi entropy is
conserved during a quantum quench with a gauge-preserving
Hamiltonian \cite{Halasz}.

For a more generic Hamiltonian, the $\mathbb{Z}_2$ gauge structure
is broken. This means that the spin configurations with an odd
number of spins flipped on a boundary loop of $\partial A$ are
allowed, therefore the topological constraint is no longer connected
to the subsystem boundary $\partial A$. Note that we can also
achieve an effective gauge structure breaking by drawing the
boundary loops of $\partial A$ on the dual lattice rather than on
the real lattice (see Fig. \ref{fig-4}) because they can then
intersect with the spin loops on the dual lattice at an arbitrary
number of points. To recover the robustness of $S_2^T$ in such a
non-gauge-preserving case, one needs to calculate it by using the
original subsystem $A$ (thick subsystem). This complicates the
situation because the reduced density matrix $\rho_A$ is not
diagonal and so Eq. (\ref{eq-ord-ent-S-2}) becomes invalid. However,
we believe that if a generalization of the R\'{e}nyi entropy formula
is found, the results in this paper can be extended to the more
generic non-gauge-preserving case as well. This further step is
crucial for verifying the robustness of the topological R\'{e}nyi
entropy against generic perturbations and hence proving its
applicability as a non-local order parameter for topologically
ordered phases. \\

%%%%%%%%%%%%%%%%%%%%%%%%%%%%%%%%%%%%%%%%%%%%%%%%

\begin{acknowledgments}

We thank X.-G. Wen for illuminating discussions. This work was
supported in part by the National Basic Research Program of China
Grants No. 2011CBA00300 and No. 2011CBA00301 and the National
Natural Science Foundation of China Grants No. 61073174, No.
61033001, and No. 61061130540. Research at the Perimeter Institute
for Theoretical Physics is supported in part by the Government of
Canada through NSERC and by the Province of Ontario through MRI.

\end{acknowledgments}

%%%%%%%%%%%%%%%%%%%%%%%%%%%%%%%%%%%%%%%%%%%%%%%%

\appendix*

\section{Detailed descriptions of the PCUT calculations in the actual 2D case} \label{sec-app}

\subsection{PCUT calculation at small magnetic field}
\label{sec-app-sm}

When considering Eq. (\ref{eq-2D-H}) in the $\lambda \ll 1$ limit,
we can use the PCUT procedure to relate the eigenstates of the
perturbed Hamiltonian $\hat{H}$ with $\lambda > 0$ to those of the
unperturbed Hamiltonian $\hat{H}_0$ with $\lambda = 0$. This method
relies on the concept of elementary excitations. In the case of
$\hat{H}_0$, these excitations are flips of stars (quasi-spins)
$A_s^z$ with an energy cost of $2$ for each. They appear pairwise
when switching on the perturbation, and the perturbed Hamiltonian
can be written as
\begin{equation}
\hat{H} = 2\hat{Q} + \hat{T}_{+2} + \hat{T}_0 + \hat{T}_{-2},
\label{eq-app-sm-H-old}
\end{equation}
where $\hat{Q}$ counts the number of excitations, and $\hat{T}_n$ is
the component of the perturbation that changes the number of
excitations by $n$. It can be verified that $[\hat{Q}, \hat{T}_n] =
n \hat{T}_n$ and that $\hat{T}_n^{\dag} = \hat{T}_{-n}$. The
explicit forms of the terms in Eq. (\ref{eq-app-sm-H-old}) are
\begin{eqnarray}
\hat{Q} &=& \frac{1}{2} \sum_s \left( 1 - \hat{A}_s^z \right), \quad
\hat{T}_{+2} = - \lambda \sum_{\langle s,s' \rangle}
\hat{A}_s^{-} \hat{A}_{s'}^{-}, \label{eq-app-sm-QT} \\
\hat{T}_0 &=& - \lambda \sum_{\langle s,s' \rangle} \left(
\hat{A}_s^{+} \hat{A}_{s'}^{-} + \hat{A}_s^{-} \hat{A}_{s'}^{+}
\right), \, \, \hat{T}_{-2} = - \lambda \sum_{\langle s,s' \rangle}
\hat{A}_s^{+} \hat{A}_{s'}^{+}, \nonumber
\end{eqnarray}
where $\hat{A}_s^{\pm} = (\hat{A}_s^x \pm i \hat{A}_s^y) / 2$ are
the standard spin raising and lowering operators. In the basis of
the $\hat{H}_0$ excitations, the term $\hat{Q}$ is diagonal, while
the terms $\hat{T}_n$ are non-diagonal. The application of the PCUT
involves an iterative sequence of steps to construct a unitary basis
transformation $\hat{U}(l)$ such that the transformed Hamiltonian
$\hat{H}(l) = \hat{U}^{\dag}(l) \hat{H} \hat{U}(l)$ changes
continuously from $\hat{H}$ at $l=0$ to a block-diagonal form at $l
\rightarrow \infty$. The blocks in the asymptotic form $\hat{H}'
\equiv \hat{H}(\infty)$ correspond to subspaces of constant
excitation number, and the excitations can be found by solving the
blocks. Note that these excitations belong to the perturbed
Hamiltonian $\hat{H}$, therefore they are not the same as the
original $\hat{H}_0$ excitations. To avoid confusion, we refer to
them as quasi-excitations.

According to the standard procedure of the PCUT \cite{Dusuel, PCUT},
we write $\hat{H}(l) = 2\hat{Q} + \hat{T}_{+2}(l) + \hat{T}_0(l) +
\hat{T}_{-2}(l)$ as in Eq. (\ref{eq-app-sm-H-old}), and define
$\hat{\eta}(l) \equiv \hat{T}_{+2}(l) - \hat{T}_{-2}(l)$. If we then
require $\hat{U}(l)$ to satisfy the equation $\partial_l \hat{U}(l)
= - \hat{U}(l) \hat{\eta}(l)$, it follows that $\partial_l
\hat{H}(l) = [\hat{\eta}(l), \hat{H}(l)]$. In terms of the
components $\hat{T}_n (l)$, this equation for $\hat{H}(l)$ becomes
\begin{eqnarray}
\partial_l \hat{T}_0(l) &=& 2 \left[ \hat{T}_{+2}(l), \hat{T}_{-2}(l)
\right], \nonumber \\
\partial_l \hat{T}_{+2}(l) &=& -4 \hat{T}_{+2}(l) +
\left[ \hat{T}_{+2}(l), \hat{T}_0(l) \right], \quad
\label{eq-app-sm-UT-1} \\
\partial_l \hat{T}_{-2}(l) &=& -4 \hat{T}_{-2}(l) +
\left[ \hat{T}_0(l), \hat{T}_{-2}(l) \right]. \nonumber
\end{eqnarray}
The last two equations show that $\hat{T}_{+2}(\infty) =
\hat{T}_{-2}(\infty) = 0$, which is consistent with the
block-diagonal form of $\hat{H}(\infty)$. To solve the equations for
$\hat{U}(l)$ and $\hat{T}_n(l)$ iteratively, we write these
quantities in a series as
\begin{equation}
\hat{U}(l) = \sum_{k=0}^{\infty} \hat{U}^{(k)}(l), \quad
\hat{T}_n(l) = \sum_{k=1}^{\infty} \hat{T}_n^{(k)}(l).
\label{eq-app-sm-UT-2}
\end{equation}
The equations for $\hat{U}(l)$ and $\hat{T}_n(l)$ then take the
forms
\begin{eqnarray}
\partial_l \hat{U}^{(k)}(l) &=& - \sum_{j=0}^{k-1} \hat{U}^{(j)}(l)
\Big{\{} \hat{T}_{+2}^{(k-j)}(l) - \hat{T}_{-2}^{(k-j)}(l) \Big{\}},
\nonumber \\
\partial_l \hat{T}_0^{(k)}(l) &=& 2 \sum_{j=1}^{k-1} \left[
\hat{T}_{+2}^{(j)}(l), \hat{T}_{-2}^{(k-j)}(l) \right],
\label{eq-app-sm-UT-3} \\
\partial_l \hat{T}_{+2}^{(k)}(l) &=& -4 \hat{T}_{+2}^{(k)}(l) +
\sum_{j=1}^{k-1} \left[ \hat{T}_{+2}^{(j)}(l), \hat{T}_0^{(k-j)}(l)
\right], \nonumber \\
\partial_l \hat{T}_{-2}^{(k)}(l) &=& -4 \hat{T}_{-2}^{(k)}(l) +
\sum_{j=1}^{k-1} \left[ \hat{T}_0^{(j)}(l), \hat{T}_{-2}^{(k-j)}(l)
\right], \nonumber
\end{eqnarray}
and the corresponding starting conditions at $l = 0$ become
\begin{eqnarray}
\hat{U}^{(k)}(0) &=& \Big{\{} \begin{array}{c} 1 \qquad (k = 0) \\
0 \qquad (k \geq 1) \end{array}, \label{eq-app-sm-UT-4} \\
\hat{T}_n^{(k)}(0) &=& \Big{\{} \begin{array}{c} \hat{T}_n \qquad \, (k = 1) \\
0 \qquad \quad (k \geq 2) \end{array}. \nonumber
\end{eqnarray}
If we apply the PCUT up to second order in $\lambda$, the relevant
terms in the series of Eq. (\ref{eq-app-sm-UT-2}) are
\begin{eqnarray}
\hat{U}^{(0)}(l) &=& 1, \quad \hat{U}^{(1)}(l) = -\frac{1}{4} \left(
\hat{T}_{+2} - \hat{T}_{-2} \right) \left( 1 - e^{-4l}
\right), \nonumber \\
\hat{U}^{(2)}(l) &=& \frac{1}{32} \left( \hat{T}_{+2} - \hat{T}_{-2}
\right)^2 \left( 1 - e^{-4l} \right)^2
\label{eq-app-sm-UT-5} \\
&-& \frac{1}{16} \left[ \hat{T}_{+2} + \hat{T}_{-2} \, , \,
\hat{T}_0 \right] \Big{(} 1 - (1 + 4l) \, e^{-4l} \Big{)}, \nonumber \\
\hat{T}_0^{(1)}(l) &=& \hat{T}_0, \quad \hat{T}_0^{(2)}(l) =
\frac{1}{4} \left[ \hat{T}_{+2}, \hat{T}_{-2} \right] \left( 1 -
e^{-8l} \right), \nonumber \\
\hat{T}_{\pm 2}^{(1)}(l) &=& \hat{T}_{\pm 2} \, e^{-4l}, \quad
\hat{T}_{\pm 2}^{(2)}(l) = \pm \left[ \hat{T}_{\pm 2}, \hat{T}_0
\right] l \, e^{-4l}, \nonumber
\end{eqnarray}
the basis transformation $\hat{U} \equiv \hat{U}(\infty)$ becomes
\begin{eqnarray}
\hat{U} &=& 1 + \frac{1}{4} \left( \hat{T}_{-2} - \hat{T}_{+2}
\right) + \frac{1}{16} \left( \left[ \hat{T}_0, \hat{T}_{-2}
\right] - \left[ \hat{T}_{+2}, \hat{T}_0 \right] \right) \nonumber \\
&+& \frac{1}{32} \left( \hat{T}_{+2} \hat{T}_{+2} + \hat{T}_{-2}
\hat{T}_{-2} - \hat{T}_{+2} \hat{T}_{-2} - \hat{T}_{-2} \hat{T}_{+2}
\right), \nonumber \\
\label{eq-app-sm-U}
\end{eqnarray}
and the asymptotic Hamiltonian takes the form
\begin{equation}
\hat{H}' = 2\hat{Q} + \hat{T}_0 + \frac{1}{4} \left[ \hat{T}_{+2},
\hat{T}_{-2} \right], \label{eq-app-sm-H-new}
\end{equation}
which indeed conserves the number of quasi-excitations. The same
procedure can be continued to arbitrary order in $\lambda$, but the
calculations quickly become cumbersome.

Since the ground state $| \Omega (\lambda) \rangle$ of $\hat{H}$ is
the only state with no quasi-excitations, it has its own block in
$\hat{H}'$. To express this state in terms of the physically
transparent $\hat{H}_0$ excitations, we use the basis
transformation: $| \Omega (\lambda) \rangle = \hat{U} | 0 \rangle$.
When calculating the first two perturbative corrections to the
ground state at $\lambda = 0$, the perturbed state $| \Omega
(\lambda) \rangle$ needs to be properly normalized up to
$\lambda^4$. Applying the PCUT up to fourth order with the aid of a
computer, the perturbed ground state becomes
\begin{eqnarray}
| \Omega (\lambda) \rangle &=& \hat{U} | 0 \rangle = \left[ 1 -
\frac{N^2 \lambda^2}{16} + \frac{(N^4 - 95N^2) \lambda^4}{512}
\right] | 0 \rangle \nonumber \\
&+& \left[ \frac{\lambda}{4} - \frac{(N^2 - 15) \lambda^3}{64}
\right] \sum_{2N^2} | \begin{array}{cc} \times & \times
\end{array} \rangle \label{eq-app-sm-gs} \\
&+& \frac{\lambda^2}{4} \sum_{2N^2} \bigg{|} \begin{array}{cc}
\times & \cdot \\ \cdot & \times \end{array} \bigg{\rangle} +
\frac{\lambda^2}{8} \sum_{2N^2} | \begin{array}{ccc} \times &
\cdot & \times \end{array} \rangle \nonumber \\
&+& \frac{\lambda^2}{8} \sum_{N^2} \bigg{|} \begin{array}{cc} \times
& \times \\ \times & \times \end{array} \bigg{\rangle} +
\frac{\lambda^2}{16} \sum_{2N^4 - 9N^2} \bigg{|} \bigg{[}
\begin{array}{c} \times \\ \times \end{array} \bigg{]} \bigg{[}
\begin{array}{c} \times \\ \times \end{array} \bigg{]} \bigg{\rangle},
\nonumber
\end{eqnarray}
where the equivalent states related to each other by translational
and rotational symmetries are labeled by the relative positions of
the star excitations ($\times$) in them, and the number of states in
each equivalence class is given by the number below the
corresponding sum. The notation $[ \ldots ][ \ldots ]$ means that
there are two clusters of excitations that are independent of each
other: they are not in a relative position characterizing any other
equivalence class.

The ground state in Eq. (\ref{eq-app-sm-gs}) is indeed properly
normalized up to fourth order in $\lambda$ because
\begin{eqnarray}
\langle \Omega (\lambda) | \Omega (\lambda) \rangle &=& \left[ 1 -
\frac{N^2 \lambda^2}{16} + \frac{(N^4 - 95N^2) \lambda^4}{512}
\right]^2 \nonumber \\
&+& 2N^2 \left[ \frac{\lambda}{4} - \frac{(N^2 - 15) \lambda^3}{64}
\right]^2 \label{eq-app-sm-norm} \\
&+& 2N^2 \left( \frac{\lambda^2}{4} \right)^2 + \left( 2N^2 + N^2
\right) \left( \frac{\lambda^2}{8} \right)^2 \nonumber \\
&+& \left( 2N^4 - 9N^2 \right) \left( \frac{\lambda^2}{16} \right)^2
= 1 + O(\lambda^6). \nonumber
\end{eqnarray}
To calculate the R\'{e}nyi entropy, we need to evaluate the
expectation values of the products appearing in Eq.
(\ref{eq-ord-ent-S-2}) for the ground state. The expectation values
having a contribution up to $\lambda^4$ to the R\'{e}nyi entropy are
\begin{eqnarray}
\langle \begin{array}{cc} \times & \times \end{array} \rangle &=& 2
\left[ \frac{\lambda}{4} - \frac{(N^2 - 15) \lambda^3}{64} \right]
\nonumber \\
&\times& \Bigg{[} \left[ 1 - \frac{N^2 \lambda^2}{16} \right] + 4
\left( \frac{\lambda^2}{4} \right) \nonumber \\
&+& 4 \left( \frac{\lambda^2}{8} \right) + \left( 2N^2 - 9 \right)
\left( \frac{\lambda^2}{16} \right) \Bigg{]} \nonumber \\
&=& \frac{\lambda}{2} + \frac{15\lambda^3}{16} + O(\lambda^5),
\label{eq-app-sm-exp} \\
\bigg{\langle} \begin{array}{cc} \times & \cdot \\ \cdot & \times
\end{array} \bigg{\rangle} &=& 2 \left( \frac{\lambda^2}{4} \right)
+ 4 \left[ \frac{\lambda}{4} \right]^2 = \frac{3\lambda^2}{4} +
O(\lambda^4), \nonumber \\
\langle \begin{array}{ccc} \times & \cdot & \times \end{array}
\rangle &=& 2 \left( \frac{\lambda^2}{8} \right) + 2 \left[
\frac{\lambda}{4} \right]^2 = \frac{3\lambda^2}{8} +
O(\lambda^4), \nonumber \\
\bigg{\langle} \begin{array}{cc} \times & \times \\ \times & \times
\end{array} \bigg{\rangle} &=& 2 \left( \frac{\lambda^2}{8} \right)
+ 4 \left[ \frac{\lambda}{4} \right]^2 = \frac{\lambda^2}{2} +
O(\lambda^4), \nonumber \\
\bigg{\langle} \bigg{[} \begin{array}{c} \times \\ \times
\end{array} \bigg{]} \bigg{[} \begin{array}{c} \times \\ \times
\end{array} \bigg{]} \bigg{\rangle} &=& 2 \left( \frac{\lambda^2}{16}
\right) + 2 \left[ \frac{\lambda}{4} \right]^2 = \frac{\lambda^2}{4}
+ O(\lambda^4), \nonumber
\end{eqnarray}
where the notation is analogous to that in Eq. (\ref{eq-app-sm-gs}).
For example, $\langle \begin{array}{cc} \times & \times
\end{array} \rangle \equiv \langle \Omega (\lambda) | \hat{A}_s^x
\hat{A}_{s'}^x | \Omega (\lambda) \rangle$, where $s$ and $s'$ are
any two nearest-neighbor stars. If the boundary $\partial A$
consists of $n$ closed loops with a combined length $L$ and a total
number of $K$ corners that are sufficiently far away from each
other, the R\'{e}nyi entropy is given by
\begin{eqnarray}
S_2^{\partial A} &=& (L - n) - \log_2 \Bigg{[} 1 + L \, \langle
\begin{array}{cc} \times & \times \end{array} \rangle^2 + K \,
\bigg{\langle} \begin{array}{cc} \times & \cdot \\ \cdot & \times
\end{array} \bigg{\rangle}^2 \nonumber \\
&+& (L-K) \langle \begin{array}{ccc} \times & \cdot & \times
\end{array} \rangle^2 + \frac{L(L-3)}{2} \bigg{\langle} \bigg{[}
\begin{array}{c} \times \\ \times \end{array} \bigg{]} \bigg{[}
\begin{array}{c} \times \\ \times \end{array} \bigg{]}
\bigg{\rangle}^2 \Bigg{]} \nonumber \\
&=& (L - n) - \frac{1}{\ln 2} \left[ \frac{L}{4} \lambda^2 +
\frac{63L + 27K}{64} \lambda^4 + O(\lambda^6) \right]. \nonumber \\
\label{eq-app-sm-S-1}
\end{eqnarray}
The perturbative corrections are linearly proportional to either $L$
or $K$, and they are independent of $n$.

Now we consider a Wilson loop for a square region $R$ with boundary
length $L = 4D$ and corner number $K = 4$. According to Eq.
(\ref{eq-ord-gen-W}), the Wilson loop is $W_R = 1$ for the
unperturbed ground state $| 0 \rangle$ because $A_s^z = +1$ for all
$s$. The structure of the perturbed ground state $| \Omega (\lambda)
\rangle$ in Eq. (\ref{eq-app-sm-gs}) shows that $W_R$ can only be
$-1$ instead of $+1$ if an odd number of excitations are inside $R$.
Taking into account all possibilities up to fourth order in
$\lambda$, the Wilson loop becomes
\begin{eqnarray}
W_R &=& 1 - 2 \Bigg{[} L \left[ \frac{\lambda}{4} - \frac{(N^2 - 15)
\lambda^3}{64} \right]^2 + (2L - K) \left( \frac{\lambda^2}{4}
\right)^2 \nonumber \\
&+& 2L \left( \frac{\lambda^2}{8} \right)^2 + K \left(
\frac{\lambda^2}{8} \right)^2 + L \, (2N^2 - L - 6)
\left( \frac{\lambda^2}{16} \right)^2 \Bigg{]} \nonumber \\
&=& 1 - \frac{L}{8} \, \lambda^2 - \left( \frac{L}{2} -
\frac{L^2}{128} - \frac{3K}{32} \right) \lambda^4 + O(\lambda^6)
\nonumber \\
&=& \exp \left[ - L \left[ \frac{\lambda^2}{8} + \frac{\lambda^4}{2}
+ O(\lambda^6) \right] + K \left[ \frac{3\lambda^4}{32} +
O(\lambda^6) \right] \right]. \nonumber \\
\label{eq-app-sm-W-1}
\end{eqnarray}
The perturbative corrections inside the exponential are linearly
proportional to either $L$ or $K$.

\begin{figure}[t!]
\centering
\includegraphics[width=7.2cm]{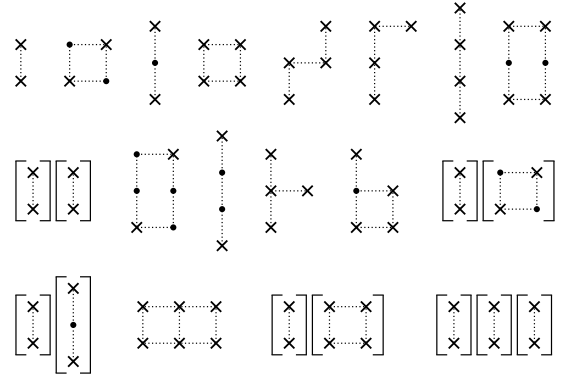}
\caption{Equivalence classes for the relative positions of the star
excitations ($\times$) at the level of the third corrections (at
order $\lambda^6$). \label{fig-7}}
\end{figure}

The expressions in Eqs. (\ref{eq-app-sm-S-1}) and
(\ref{eq-app-sm-W-1}) give the first two corrections to
$S_2^{\partial A}$ and $W_R$. With the aid of a computer, the third
corrections proportional to $\lambda^6$ can be found in a similar
manner. In this case, the state $| \Omega (\lambda) \rangle$ must be
properly normalized up to sixth order in $\lambda$, and one needs to
consider all the relative excitation positions shown in Fig.
\ref{fig-7}. Without including the detailed calculations, the final
results for the R\'{e}nyi entropy and the Wilson loop are
\begin{eqnarray}
S_2^{\partial A} &=& (L - n) - \frac{L}{\ln 2} \left[
\frac{\lambda^2}{4} + \frac{63\lambda^4}{64} +
\frac{503\lambda^6}{96} + O(\lambda^8) \right] \nonumber \\
&-& \frac{K}{\ln 2} \left[ \frac{27\lambda^4}{64} +
\frac{737\lambda^6}{256} + O(\lambda^8) \right],
\label{eq-app-sm-S-2}
\end{eqnarray}
\begin{eqnarray}
W_R &=& \exp \bigg{\{} - L \left[ \frac{\lambda^2}{8} +
\frac{\lambda^4}{2} + \frac{7697\lambda^6}{3072} + O(\lambda^8)
\right] \nonumber \\
&+& K \left[ \frac{3\lambda^4}{32} + \frac{89\lambda^6}{128} +
O(\lambda^8) \right] \bigg{\}}. \label{eq-app-sm-W-2}
\end{eqnarray}
The features noticed after the first two corrections remain intact
after the third corrections as well.

\subsection{PCUT calculation at large magnetic field}
\label{sec-app-lg}

When considering Eq. (\ref{eq-2D-lg-H}) in the $\mu \equiv
\lambda^{-1} \ll 1$ limit, the PCUT procedure is entirely analogous
to the one described in Appendix \ref{sec-app-sm}. The elementary
excitations of the unperturbed Hamiltonian with $\mu = 0$ are flips
of physical spins $\sigma_i^z$ with an energy cost of 2 for each.
The perturbed Hamiltonian with $\mu > 0$ can be written as
\begin{equation}
\hat{H} = 2\hat{Q} + \hat{T}_{+4} + \hat{T}_{+2} + \hat{T}_0 +
\hat{T}_{-2} + \hat{T}_{-4}, \label{eq-app-lg-H}
\end{equation}
where the respective terms take the explicit forms
\begin{eqnarray}
\hat{Q} &=& \frac{1}{2} \sum_i \left( 1 - \hat{\sigma}_i^z \right),
\label{eq-app-lg-QT} \\
\hat{T}_n &=& - \mu \sum_{s} \sum_{\pm} \prod_{i \in s}
\hat{\sigma}_{i}^{\pm}. \nonumber
\end{eqnarray}
The sum in $\pm$ contains all inequivalent products of the four
$\hat{\sigma}_i^{\pm}$ operators in which the number of the
$\hat{\sigma}_i^{+}$ factors is $2 - n/2$ and that of the
$\hat{\sigma}_i^{-}$ factors is $2 + n/2$. Applying the PCUT with
$\hat{\eta}(l) \equiv \hat{T}_{+4}(l) + \hat{T}_{+2}(l) -
\hat{T}_{-2}(l) - \hat{T}_{-4}(l)$ up to fourth order, we find that
the perturbed ground state which is properly normalized up to
$\mu^4$ is given by
\begin{eqnarray}
| \Omega (\lambda) \rangle &=& \left[ 1 - \frac{N^2 \mu^2}{128} +
\left( N^4 - \frac{62N^2}{9} \right) \frac{\mu^4}{32768} \right] |
\Uparrow \, \rangle \nonumber \\
&+& \left[ \frac{\mu}{8} - \left( N^2 - \frac{2}{3} \right)
\frac{\mu^3}{1024} \right] \sum_{N^2} \Bigg{|} \begin{array}{ccc} &
\circ & \\ \circ & & \circ \\ & \circ & \end{array}
\Bigg{\rangle} \nonumber \\
&+& \frac{\mu^2}{48} \sum_{2N^2} \Bigg{|} \begin{array}{ccccc} &
\circ & & \circ & \\ \circ & & \cdot & & \circ \\ &
\circ & & \circ & \end{array} \Bigg{\rangle} \label{eq-app-lg-gs} \\
&+& \frac{\mu^2}{64} \sum_{\frac{1}{2} (N^4 - 5N^2)} \Bigg{|}
\Bigg{[} \begin{array}{ccc} & \circ & \\ \circ & & \circ \\
& \circ & \end{array} \Bigg{]} \Bigg{[} \begin{array}{ccc} & \circ &
\\ \circ & & \circ \\ & \circ & \end{array} \Bigg{]} \Bigg{\rangle},
\nonumber
\end{eqnarray}
where the equivalent states are labeled by the relative positions of
the spin excitations ($\circ$) in them [cf. Eq.
(\ref{eq-app-sm-gs})].

We now consider a subsystem with a boundary $\partial A$ containing
$L$ spins and $L$ stars acting on these spins. The total number of
corners is $K$ as above. The diagonal elements of the density matrix
$\rho_{\partial A}$ can be obtained directly in the basis of the
physical spins $\sigma_i^z$. The element corresponding to
$\sigma_i^z = +1$ for all $L$ spins (no star excitations on the
boundary) is
\begin{eqnarray}
\left( \rho_{\partial A} \right)_{00} &=& \left[ 1 - \frac{N^2
\mu^2}{128} + \left( N^4 - \frac{62N^2}{9} \right)
\frac{\mu^4}{32768} \right]^2
\nonumber \\
&+& \left( N^2 - L \right) \left[ \frac{\mu}{8} - \left( N^2 -
\frac{2}{3} \right) \frac{\mu^3}{1024} \right]^2 \nonumber \\
&+& \left( 2N^2 - 3L \right) \left( \frac{\mu^2}{48} \right)^2
\label{eq-app-lg-diag-1} \\
&+& \frac{1}{2} \left[ N^4 - N^2 \left( 2L + 5 \right) + \left( L^2
+ 7L \right) \right] \left( \frac{\mu^2}{64} \right)^2 \nonumber \\
&=& 1 - \frac{L}{64} \, \mu^2 - \left( \frac{5L}{8192} -
\frac{L^2}{8192} \right) \mu^4 + O(\mu^6), \nonumber
\end{eqnarray}
while the one corresponding to $\sigma_i^z = -1$ for any two
neighboring spins and $\sigma_i^z = +1$ for the remaining $L - 2$
spins (one star excitation on the boundary) is
\begin{equation}
\left( \rho_{\partial A} \right)_{11} = \bigg{[} \frac{\mu}{8}
\bigg{]}^2 = \frac{\mu^2}{64} + O(\mu^4). \label{eq-app-lg-diag-2}
\end{equation}
Note that there are $L$ ways of choosing two neighboring spins from
$\partial A$. Since the contribution of the remaining diagonal
elements is $O(\mu^6)$ to the R\'{e}nyi entropy, we find that
\begin{eqnarray}
\mathrm{Tr} \left[ \hat{\rho}_{\partial A}^2 \right] &=& \left(
\rho_{\partial A} \right)_{00}^2 + L \left( \rho_{\partial A}
\right)_{11}^2 \label{eq-app-lg-tr} \\
\nonumber \\
&=& 1 - \frac{L}{32} \, \mu^2 - \left( \frac{L}{1024} -
\frac{L^2}{2048} \right) \mu^4 + O(\mu^6), \nonumber \\
\nonumber
\end{eqnarray}
and the R\'{e}nyi entropy takes the form
\begin{equation}
S_2^{\partial A} = \frac{1}{\ln 2} \left[ \frac{L}{32} \, \mu^2 +
\frac{L}{1024} \, \mu^4 + O(\mu^6) \right]. \label{eq-app-lg-S-1}
\end{equation}
The perturbative corrections are again linearly proportional to the
boundary length $L$. Furthermore, there are no terms $\propto K$ in
the first two corrections determined here.

\begin{figure}[h!]
\centering
\includegraphics[width=7.2cm]{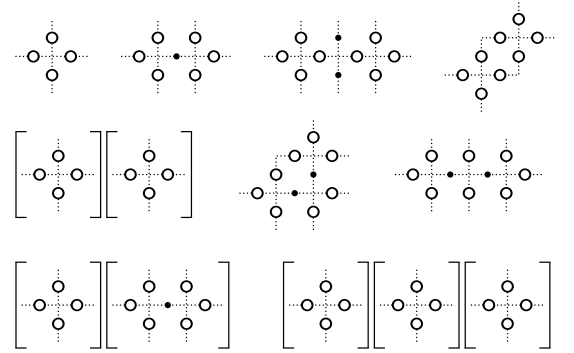}
\caption{Equivalence classes for the relative positions of the spin
excitations ($\circ$) at the level of the third corrections (at
order $\mu^6$). \label{fig-8}}
\end{figure}

With the aid of a computer, the third correction to $S_2^{\partial
A}$ can be calculated in a similar manner. In this case, the state
$| \Omega (\lambda) \rangle$ must be properly normalized up to sixth
order in $\mu$, and one needs to consider all the relative
excitation positions shown in Fig. \ref{fig-8}. Without including
the detailed calculations, the final result for the R\'{e}nyi
entropy is
\begin{eqnarray}
S_2^{\partial A} &=& \frac{L}{\ln 2} \left[ \frac{\lambda^{-2}}{32}
+ \frac{\lambda^{-4}}{1024} +
\frac{115\lambda^{-6}}{2359296} + O(\lambda^{-8}) \right] \nonumber \\
&-& \frac{K}{\ln 2} \left[ \frac{35\lambda^{-6}}{4718592} +
O(\lambda^{-8}) \right]. \label{eq-app-lg-S-2}
\end{eqnarray}
After the third correction, there is a corner contribution $\propto
K$, but still no topological contribution $\propto n$.

%%%%%%%%%%%%%%%%%%%%%%%%%%%%%%%%%%%%%%%%%%%%%%%%

%%%%%%%%%%%%%%%%%%%%%%%%%%%%%%%%%%%%%%%%%%%%%%%%


\begin{references}

\bibitem{Wen} X.-G. Wen, \emph{Quantum Field Theory of Many-Body
Systems} (Oxford University Press, Oxford, 2004).
\bibitem{FQHE} X.-G. Wen, Adv. Phys. \textbf{44}, 405 (1995); H. L.
Stormer, D. C. Tsui, and A. C. Gossard, Rev. Mod. Phys. \textbf{71},
S298 (1999).
\bibitem{Isakov} S. V. Isakov, M. B. Hastings, and R. G. Melko,
Nat. Phys. \textbf{7}, 772 (2011).
\bibitem{Kitaev} A. Y. Kitaev, Ann. Phys. (N. Y.) \textbf{303}, 2
(2003).
\bibitem{QC} M. H. Freedman, A. Kitaev, and Z. Wang, Commun. Math.
Phys. \textbf{227}, 587 (2002); C. Nayak, S. H. Simon, A. Stern, M.
Freedman, and S. Das Sarma, Rev. Mod. Phys. \textbf{80}, 1083
(2008).
\bibitem{Hamma-1} A. Hamma, R. Ionicioiu, and P. Zanardi, Phys.
Lett. A \textbf{337}, 22 (2005); A. Hamma, R. Ionicioiu, and P.
Zanardi, Phys. Rev. A \textbf{71}, 022315 (2005).
\bibitem{TE} A. Kitaev and J. Preskill, Phys. Rev. Lett.
\textbf{96}, 110404 (2006); M. Levin and X.-G. Wen, \emph{ibid.}
\textbf{96}, 110405 (2006).
\bibitem{Flammia} S. T. Flammia, A. Hamma, T. L. Hughes, and X.-G.
Wen, Phys. Rev. Lett. \textbf{103}, 261601 (2009).
\bibitem{Kim} I. H. Kim, Phys. Rev. B \textbf{86}, 245116 (2012).
\bibitem{Chen} X. Chen, Z.-C. Gu, and X.-G. Wen, Phys. Rev. B
\textbf{82}, 155138 (2010).
\bibitem{Hamma-2} A. Hamma and D. A. Lidar, Phys. Rev. Lett.
\textbf{100}, 030502 (2008).
\bibitem{Hamma-3} A. Hamma, W. Zhang, S. Haas, and D. A. Lidar,
Phys. Rev. B \textbf{77}, 155111 (2008).
\bibitem{Amico} L. Amico, R. Fazio, A. Osterloh, and V. Vedral,
Rev. Mod. Phys. \textbf{80}, 517 (2008).
\bibitem{Papanikolaou} S. Papanikolaou, K. S. Raman, and E.
Fradkin, Phys. Rev. B \textbf{76}, 224421 (2007).
\bibitem{Castelnovo} C. Castelnovo and C. Chamon, Phys. Rev. B
\textbf{77}, 054433 (2008).
\bibitem{Levin} M. A. Levin and X.-G. Wen, Phys. Rev. B
\textbf{71}, 045110 (2005).
\bibitem{Buerschaper} O. Buerschaper and M. Aguado, Phys. Rev. B
\textbf{80}, 155136 (2009).
\bibitem{Goldenfeld} N. Goldenfeld, \emph{Lectures on Phase
Transitions and the Renormalization Group} (Addison-Wesley, New
York, 1992).
\bibitem{Trebst} S. Trebst, P. Werner, M. Troyer, K. Shtengel, and
C. Nayak, Phys. Rev. Lett. \textbf{98}, 070602 (2007).
\bibitem{Dusuel} S. Dusuel, M. Kamfor, K. P. Schmidt, R. Thomale,
and J. Vidal, Phys. Rev. B \textbf{81}, 064412 (2010).
\bibitem{Yu} J. Yu, S.-P. Kou, and X.-G. Wen, Europhys. Lett.
\textbf{84}, 17004 (2008).
\bibitem{Barouch-1} E. Barouch, B. M. McCoy, and M. Dresden, Phys.
Rev. A \textbf{2}, 1075 (1970).
\bibitem{Barouch-2} E. Barouch and B. M. McCoy, Phys. Rev. A
\textbf{3}, 786 (1971).
\bibitem{PCUT} F. Wegner, Ann. Phys. (Leipzig) \textbf{3}, 77
(1994); S. D. G{\l}azek and K. G. Wilson, Phys. Rev. D \textbf{48},
5863 (1993); S. D. G{\l}azek and K. G. Wilson, \emph{ibid.}
\textbf{49}, 4214 (1994); S. Dusuel and J. Vidal, Phys. Rev. B
\textbf{71}, 224420 (2005); S. Dusuel, M. Kamfor, R. Or\'{u}s, K. P.
Schmidt, and J. Vidal, Phys. Rev. Lett. \textbf{106}, 107203 (2011).
\bibitem{Ising} Q. Jiang and X.-F. Jiang, Phys. Lett. A
\textbf{224}, 196 (1997); M. S. L. du Croo de Jongh and J. M. J. van
Leeuwen, Phys. Rev. B \textbf{57}, 8494 (1998); H. Rieger and N.
Kawashima, Eur. Phys. J. B \textbf{9}, 233 (1999); H. W. J.
Bl\"{o}te and Y. Deng, Phys. Rev. E \textbf{66}, 066110 (2002).
\bibitem{Tsomokos} D. I. Tsomokos, A. Hamma, W. Zhang, S. Haas, and
R. Fazio, Phys. Rev. A \textbf{80}, 060302(R) (2009).
\bibitem{Halasz} G. B. Hal\'{a}sz and A. Hamma, arXiv:1211.5381.

\end{references}
\end{document}